\documentclass[11pt]{prop_axis} 

\usepackage[super,comma]{natbib} 
\usepackage{apjfonts} 
\usepackage{graphicx} 
\usepackage{amssymb,amsmath} 
\usepackage{color} 
\usepackage{enumitem} 
\usepackage{wrapfig} 
\usepackage{tikz} 

\usepackage{multirow} 
\usepackage{tabularx} 
\usepackage{array} 

\usepackage{hyperref} 
\hypersetup{
    colorlinks,
    linkcolor={blue!80!black},
    citecolor={blue!80!black},
    urlcolor={blue!80!black}
}
\usepackage{sidecap}
\sidecaptionvpos{figure}{c} 

\newcolumntype{L}[1]{>{\raggedright\let\newline\\\arraybackslash\hspace{0pt}}p{#1}}
\newcolumntype{C}[1]{>{\centering\let\newline\\\arraybackslash\hspace{0pt}}p{#1}}
\newcolumntype{R}[1]{>{\raggedleft\let\newline\\\arraybackslash\hspace{0pt}}p{#1}}
\renewcommand\arraystretch{1.3}

\definecolor{headcolor}{rgb}{0.65,0.65,0.65}
\usepackage{fancyhdr}
\renewcommand{\topmargin}{-9mm}

\renewcommand{\textfloatsep}{4mm}

\newcommand{\lem}{\textit{LEM}}
\newcommand{\LEM}{\textit{LEM}}

\newcommand{\XMM}{\textit{XMM-Newton}}

\newcommand{\JWST}{\textit{JWST}}

\newcommand{\ROSAT}{\textit{ROSAT}}

\newcommand{\bsf}{\sffamily\bfseries}

\definecolor{callout}{rgb}{0.25,0.40,0.85}
\definecolor{synergies}{rgb}{0.20,0.45,0.99}
\definecolor{methods}{rgb}{0.20,0.70,0.45}

\definecolor{calllem}{rgb}{0.20,0.45,0.99}
\definecolor{tabledef}{rgb}{0.95,0.95,0.95}
\definecolor{tablealt}{rgb}{0.77,0.80,1.0}
\definecolor{tablelem}{rgb}{0.80,0.85,1.0}
\definecolor{whitelem}{rgb}{1.0,1.0,1.0}
\definecolor{greenlem}{rgb}{0.7,1.0,0.7}

\begin{document}

\baselineskip=13.2pt
\sloppy
\pagenumbering{roman}
\thispagestyle{empty}

\title{\sf\huge {\sl Active Galaxy Science with the Line Emission Mapper: The Case for High-Resolution Soft X-ray Spectroscopy\footnote{Corresponding author: Kimberly Weaver (Kimberly.A.Weaver@nasa.gov)}}}

\maketitle
\vspace*{-10mm}

\begin{tikzpicture}[remember picture,overlay]
\node[anchor=north west,yshift=2pt,xshift=2pt]%
    at (current page.north west)
    {\includegraphics[height=20mm]{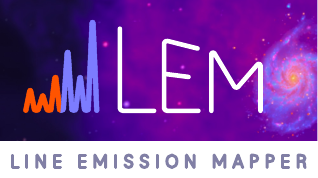}};
\end{tikzpicture}


\renewcommand{\thefootnote}{\arabic{footnote}}
\vspace*{-5mm}
\begin{center}
\begin{minipage}{16.5cm}
\hspace{-5mm}
\centering
Kimberly A. Weaver\footnotemark[1]\footnotetext{$^1$NASA Goddard Space Flight Center, Greenbelt, MD 20771, USA}, Jenna M. Cann\footnotemark[2]\footnotetext{$^2$NASA Goddard Space Flight Center, Oak Ridge Associated Universities, NASA NPP Program, Oak Ridge, TN 37831, USA}, Ryan W. Pfeifle\footnotemark[2], Malgorzata Sobolewska\footnotemark[3]\footnotetext{$^{3}$Center for Astrophysics $|$ Harvard \& Smithsonian, 60 Garden Street, Cambridge, MA 02138}, Ciro Pinto\footnotemark[4]\footnotetext{$^{4}$INAF – IASF Palermo, Via U. La Malfa 153, I-90146 Palermo, Italy}, Mojegan Azadi\footnotemark[3], Delphine Porquet\footnotemark[5]\footnotetext{$^5$Aix Marseille Univ, CNRS, CNES, LAM, Marseille, France}, Priyanka Chakraborty\footnotemark[3], Daniele Rogantini\footnotemark[6]\footnotetext{$^{6}$Department of Astronomy and Astrophysics, University of Chicago, Chicago, IL 60637, USA}, Gerrit Schellenberger\footnotemark[3], Ryan Tanner\footnotemark[7]\footnotetext{$^{7}$Catholic University of America, Washington DC, USA}, Simona Mei\footnotemark[8]\footnotetext{$^{8}$Universit\'e Paris Cit\'e, CNRS(/IN2P3), Astroparticule et Cosmologie, F-75013 Paris, France }, Akos Bogdan\footnotemark[3], and Dustin Nguyen\footnotemark[9]\footnotetext{$^{9}$Ohio State University, 281 W Lane Ave, Columbus, OH 43210}

\end{minipage}
\end{center}

\vfill

\noindent

\noindent
\phantom{${^52}$}~\textcolor{blue}{\bsf lem-observatory.org}\\

\phantom{${^52}$}~\textcolor{blue}{\bsf X / twitter: \href{https://www.twitter.com/LEMXray}{LEMXray}}\\
\phantom{${^52}$}~\textcolor{blue}{\bsf facebook: \href{https://www.facebook.com/LEMXrayProbe}{LEMXrayProbe}}\\

\centerline{\lem\ AGN white paper, 2024}
\clearpage
\twocolumn


\setcounter{page}{1}
\pagenumbering{arabic}

\section{SUMMARY}
\label{sec:summary}

This white paper discusses the breadth of science related to active galactic nuclei (AGN) and associated phenomena to be enabled by a mission with microcalorimeter energy resolution in the soft X-ray band, a large collecting area, and wide-field imaging.\ Such a mission, the {\it Line Emission Mapper}\cite{kraft2022} (\lem), has been proposed to NASA's 2023 Astrophysics Probe Explorer call.\ While the science pillars of the PI-led part of the mission focus on galaxy evolution, the PI-led \lem\ All-Sky Survey (LASS) and General Observer/Investigator opportunities will enable vital discoveries for AGN science in the critical soft X-ray band.\

\section{INTRODUCTION}
\label{sec:intro}

The {\it Line Emission Mapper}\cite{kraft2022} (\lem) combines high-resolution X-ray spectroscopy and a large collecting area with wide-field imaging in the soft X-ray band (0.2--2.0\,keV).\ The instrument consists of a large-area X-ray mirror with a baseline requirement of $18\,''$ half-power diameter, and an X-ray Integral Field Unit microcalorimeter array.\ This advanced calorimeter array covers a $30\,'$ by $30\,'$ field of view with $15\,''$ pixels ($\sim1$\,eV or $\sim2$\,eV energy resolution; E/$\Delta$E = 200 to 2000).\ \lem\ is much more sensitive than current grating instruments, and with a baseline effective area of 1,200 cm$^{2}$, will improve measurements by a factor of 10 to 100 over the \XMM\ reflection grating spectrometer (RGS) and the {\it Chandra} low and high-energy transmission gratings (LETG/HETG).\ \lem \ also covers the low energy band not currently available with the XRISM Resolve calorimeter; the X-ray aperture door has not yet opened at the time of this writing, thereby shifting Resolve's energy band from 0.3 --12 keV to 1.7--12 keV.\ With the capability to centroid emission lines to within a 50 km\,s$^{-1}$ precision, \lem\ will rival measurements in the IR, optical, and UV bands.\ Its suite of capabilities will break open science inaccessible to current dispersive spectrometers or CCDs, and allow specifically for direct comparison to ground-based and space-based missions in discrete multi-wavelength studies.

The \lem\ mission concept was created in response to the priorities outlined in the Astro2020 Decadal report\cite{astro2020}, and is aimed at unraveling the mysteries of galaxy evolution.\ There is a tremendous depth of science within the area of galaxy evolution to be covered by \lem\, such as the detailed studies of the circumgalactic medium (CGM)\cite{ZuHone2023,Bogdan2023,Schellenberger2023} and the intracluster medium (ICM)\cite{Mernier2023,Zhang2023}.\ Galaxies themselves are governed by the competition between gravity and energetic `feedback' that results from star formation and supermassive black holes (SMBH) \cite{2012ARA&A..50..455F, kingpounds15, manbelli18}.\  In this way SMBH at the hearts of AGN serve as foundational kinematic and energetic catalysts of the phenomena that make up the \lem\ PI-led science pillar that includes CGM studies\cite{kraft2022}.

In addition to this science pillar, \lem\ observations of AGN will support the Astro2020 themes of Cosmic Ecosystems and New Messengers and New Physics, with priority science areas (respectively) of Unveiling the Hidden drivers of galaxy growth and opening New Windows on the Dynamic Universe through the goal of Understanding black hole accretion and feedback\cite{astro2020}.\ The astrophysical questions of this decade posed in these areas are: (1) {\it How do gas, metals, and dust flow into, through, and out of galaxies?}, (2) {\it How do supermassive black holes form?}, and (3) {\it How is their growth coupled to the evolution of their host galaxies?}

\begin{figure*}[ht!]
    \includegraphics[width=0.99\textwidth]{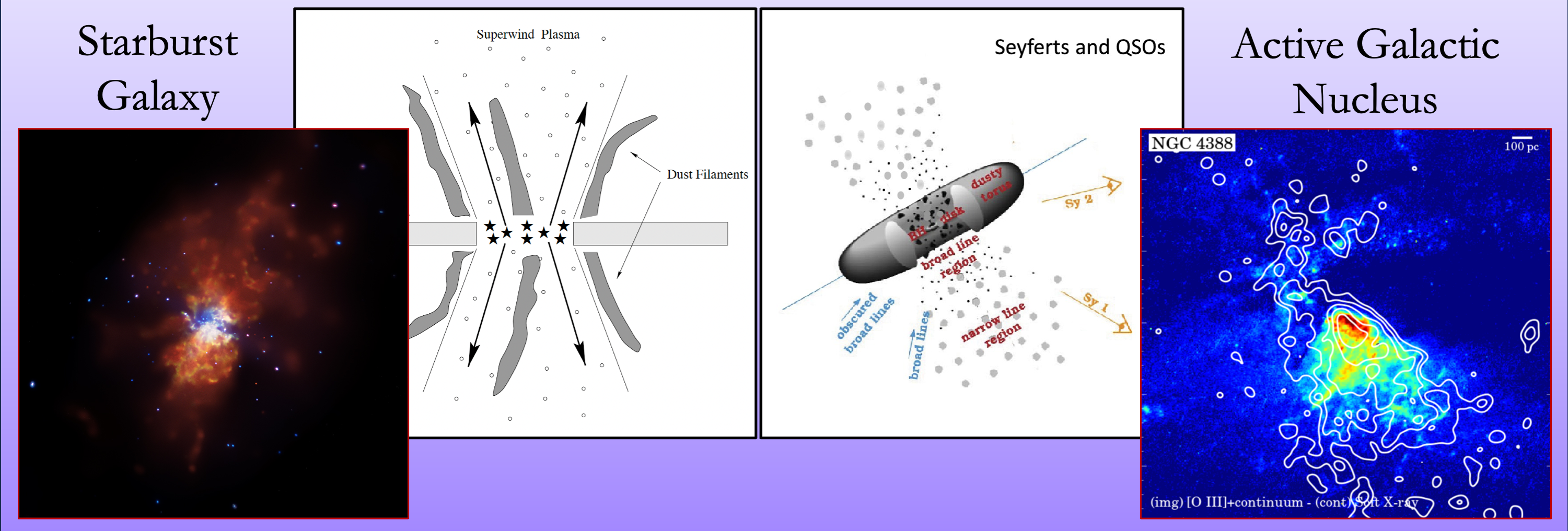}
    \caption{\footnotesize{{\bf The interplay of stellar and AGN feedback - 
\LEM\ will examine the role of AGN as catalysts in driving galaxy evolution.}\ \LEM\ will address key questions such as how AGN impact their host galaxies and how might AGN outflows drive/quench star formation.\ Left pair is the {\it Chandra} image of the starburst galaxy M82 and a model that describes the starburst outflow\cite{Alton99}.\ Right pair is a model for the geometry and energetics of a radio-quiet AGN\cite{Dermer16} and an optical image plus Chandra contours for the Type\,2 AGN NGC\,4388\cite{gomezguijarro17}.\ The soft X-ray emission aligns with high-excitation optical emission (Hubble Space Telescope, HST) that denotes the gas in the narrow line region photoionized by the activity associated with the central SMBH.}}
    \label{fig:purple}
\end{figure*}

In the local Universe, \lem\ will reveal how SMBH drive galaxy evolution by tracing AGN energy input into the interstellar medium.\ \lem\ will untangle stellar and black-hole feedback.\ For bright AGN, \lem\ will study spectral and temporal characteristics to determine outflow driving mechanisms and will provide reverberation mapping on many scales to directly uncover the accretion disk-wind connection and SMBH-galaxy connection.

\lem\ will study high-$z$ AGN, where the nearly ubiquitous Fe\,K$\alpha$ line is redshifted into the \lem\ energy band for $z$ $\gtrsim$ 2.5, i.e., covering the period of peak star formation rate in the Universe (also known as "cosmic noon").\ PI-led, pointed deep fields will reach 0.2--2 keV point source flux levels of $\sim$2--3$\times$10$^{-16}$ erg\,cm$^{-2}$\,s$^{-1}$ and will have a projected number of greater than 100 Fe K$\alpha$ lines per square degree down to flux limits of $\sim9\times10^{-17}$ erg\,cm${^{-2}}$\,s${^{-1}}$ (depending on metal abundance).\ This sensitivity probes luminosities up to L$_{2-10\,\rm{keV}}$ = $10^{47}$ erg\,s$^{-1}$ for very heavily obscured AGN. 

Here we explore the following areas of AGN discovery space to be enabled by General Observer (GO) programs, and serendipitous science that can be funded through the General Investigator (GI) program using data from the planned PI-led LASS and deep observations of PI-led targets:\
\vspace*{1.5mm}

\noindent $\bullet$ AGN and host galaxy connections 

\noindent $\bullet$ Photoionized plasmas and soft X-ray diagnostics 

\noindent $\bullet$ Black hole accretion and feedback on small scales 

\noindent $\bullet$ Properties of extreme AGN winds

\noindent $\bullet$ Large scale outflows and feedback 

\noindent $\bullet$ X-ray studies comparable to optical/UV bands

\noindent $\bullet$ Detecting AGN to high redshifts from deep fields

\noindent $\bullet$ AGN population studies at $z$ $\ge$ 2.5

\noindent $\bullet$ Fe K line properties in quasars/QSOs at $z$ $\sim$3--5

\noindent $\bullet$ AGN in protoclusters

\noindent $\bullet$ Synergies with gravitational wave facilities

\vspace*{1.5mm}
\section{AGN science with \LEM}
\label{sec:sci}
\lem\ has the power to revolutionize our understanding of AGN by providing exceptional spectra at low X-ray energies, opening the door to scientific discoveries inaccessible to CCD imaging alone.\ With its $\sim$100 km s$^{-1}$ resolution, \lem\ can compare directly with optical spectroscopy.\ By mapping gas velocities to high precision and probing gas thermodynamics, \lem\ will untangle the impact of AGN on star formation (SF) -- an effect known as `AGN feedback'.\ By detecting and separating features of AGN photoionization from thermal plasmas and shock heating from stellar sources, \lem\ will spectrally decouple AGN signals from their host galaxies.\ Black hole accretion and feedback can be measured on small scales via extreme AGN winds, and \lem\ will unveil the disk-wind connection for highly-accreting SMBH by placing constraints on outflow and net accretion rates.\ AGN feedback will also be probed on large scales for nearby galaxies where \lem\ can spatially resolve and directly measure variations such as velocity structures.\ 

At higher redshifts, \lem\ will detect tens of thousands of AGN at $z\ge2.5$ down to a limiting flux of 3$\times$10$^{-16}$\,erg\,cm$^{-2}$\,s$^{-1}$ in the 0.5 to 2\,keV band, and \lem\ will provide some of the first robust constraints for log(N)--log(S) out to $z = 10$ across its mission lifetime.\ \lem\ studies of AGN feedback will also extend to high redshifts, and \lem\ will uncover detailed Fe K line properties in quasars at z$\sim$3--5.\ A key \lem\ science goal is to study the intergalactic medium in protoclusters, and deep observations of these sources will provide serendipitous AGN science.\ To round out its impact, \lem\ will play a key role in multi-messenger astronomy by searching for X-ray counterparts of gravitational wave sources.\ Here we expand upon all of these science cases.

\begin{figure*}[h!]
    \includegraphics[width=0.99\textwidth]{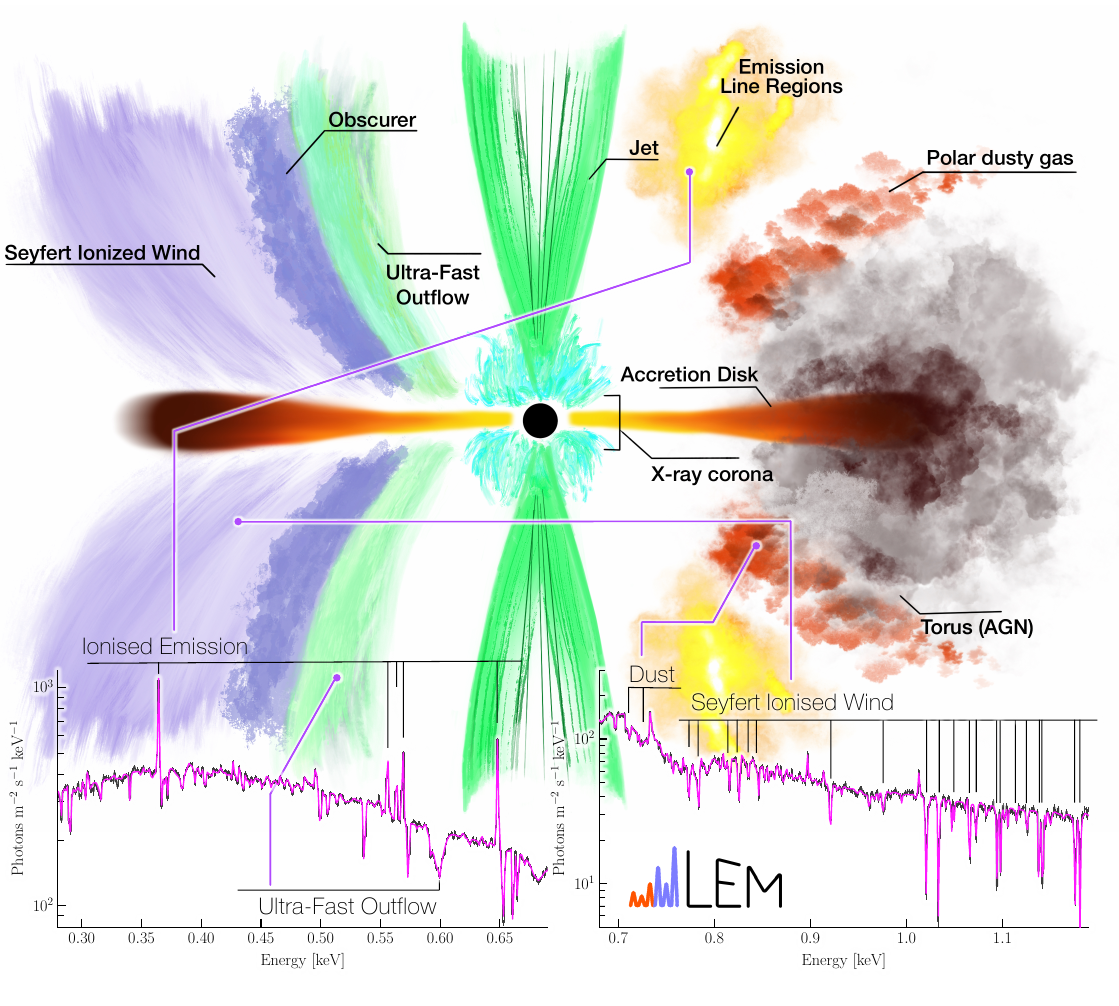}
    \caption{\footnotesize{{\bf The influence of supermassive black holes on many scales.} 
This illustration shows the complex inner workings of an AGN, with the accretion disk, ionized disk winds, and dusty torus clouds drawn.\ Along the bottom is a synthetic \lem\ spectrum of a standard local, bright Seyfert\,1 galaxy with a 50\,ks exposure time.\ The X-ray continuum of Mrk\,335 is adopted here and absorbed by a two-component Seyfert\,1 wind with an ionization parameter ($\xi=L_{\rm ion}/n\,R^{2}$ expressed in erg s$^{-1}$ cm) in logarithmic scale of 1.5 and 3, a relative ionized sub-relativistic wind and intrinsic cosmic dust close to the torus.\ We also include the narrow emission lines from a two-zone photoionized plasma with log $\xi$ of 1.1 and 2.0.}
}
    \label{fig:cartoon}
\end{figure*}

\subsection{AGN and host galaxy connections in the local Universe}
\label{sec:agnhost}

Understanding `AGN feedback' is crucial to understanding how the present-day Universe came to be.\ Significant evidence exists of correlations between SMBH mass and key properties of the host-galaxy suggesting co-evolution (e.g., bulge stellar mass\cite{kormandyho13}; bulge luminosity\cite{kormandyrichstone95, mcluredunlop02, sani11}; bulge stellar velocity dispersion\cite{ferraresemerrit00, gebhardt00, mcconnellma13}).\ However,
the specific mechanisms whereby a SMBH can influence gas and SF on size scales billions of times greater than itself remain mysterious despite years of effort\cite{2012ARA&A..50..455F, kingpounds15, manbelli18, costa18}.\ The galaxies in which feedback is operating can be highly complex ({\bf Figure~\ref{fig:purple}}).\ As we discuss below, \lem\ will allow us to disentangle the AGN from its host galaxy thanks to highly-resolved X-ray spectroscopy, coupled with spatially-resolved spectra for local AGN, and will clarify whether or to what degree feedback is controlled by direct radiation from the AGN accretion disk, from winds, or other mechanisms ({\bf Figure~\ref{fig:cartoon}}).

\begin{figure*}[ht]
    \centering
    \includegraphics[width=0.98\textwidth]{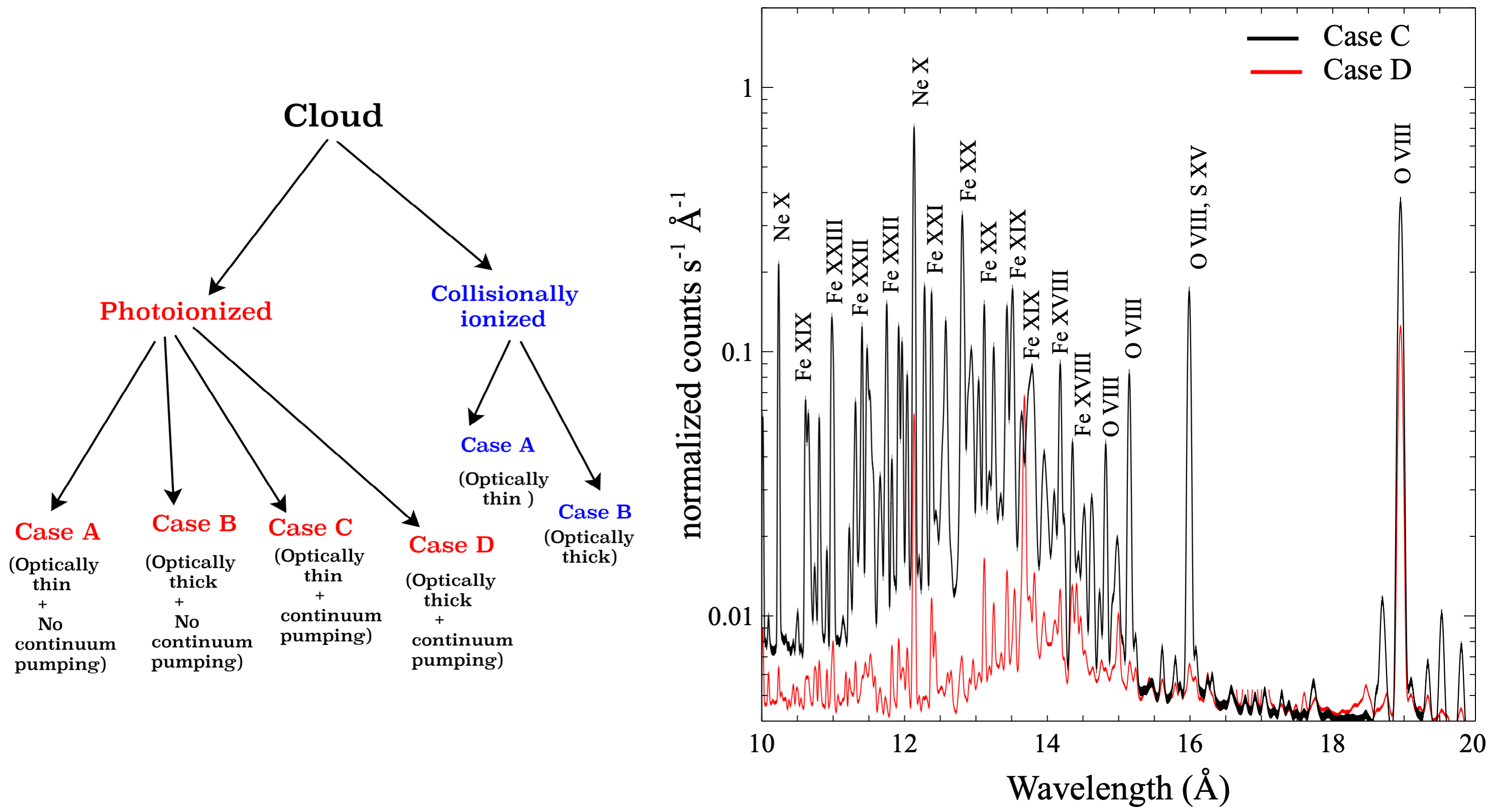}
    \caption{\footnotesize{{\bf \lem\ will make tremendous progress in studying photoionized plasmas.}\ Left: Flowchart taken from \cite{chakraborty21} showing line formation conditions in photoionized and collisionally ionized clouds.\ Right: A model comparison between Case\,C and Case\,D in AGN for $U$=1 using a power-law spectral energy distribution simulated with {\sc Cloudy} for a  30\,ks \lem\ exposure. An electron density of log $n_e$ = 10$^{10}$ cm$^{-3}$ was assumed.\ Case C and Case D simulations were performed with hydrogen column densities of log\,N$_{H}$= 21 cm$^{-2}$ and log\,N$_{H}$= 24 cm$^{-2}$, respectively. }}
    \label{fig:caseABCD}
\end{figure*}

\subsubsection{Photoionized Plasmas.} 
\label{sec:photo}

Spectral diagnostics using soft X-ray emission lines, based for example on He-like ion lines, are a powerful tool to determine the physical conditions of astrophysical plasmas, such as the ionization process, electron temperature, and the electronic density\cite{porquet10,liedahl99}.\ The process of line formation in an astrophysical plasma occurs under several different conditions and has been primarily (historically) a focus of IR, optical, and UV observational studies.\ In a low density ($n_e < 100$ cm$^{-3}$) ionized cloud, the gas is either optically thin to Lyman line radiation (low column density - Case A) or the conditions are such that the Lyman lines are optically thick, and multiple photon scatterings produce Balmer and L$\alpha$ emission lines (high column density - Case B, e.g, \cite{bakermenzel83}).\ Case A and B occur in both collisionally-ionized and photoionized clouds (see {\bf Figure~\ref{fig:caseABCD}}).\ 

Most appropriate for extragalactic sources are Cases C and D, which occur only for photoionized clouds.\ Case C relates to the power-law (simplest case) continuum source striking a cloud that is optically thin (low column density) with line emission being enhanced by radiative excitation, also called continuum pumping (emphasized in the literature by \cite{ferland99}).\ Case D occurs when the spectrum produced via Case B (high column density) is enhanced by continuum pumping - the optically-thick case for extragalactic sources \citep{luridiana}.  

\lem\ will have the capability to differentiate specifically between a Case\,C and Case\,D plasma.\ The interplay between the increase in column density, resulting in the optical depth effects \cite{chakraborty20c}, and the process of radiative cascades triggered by the absorption of line photons from an external radiation source (continuum pumping) determines the degree of enhancement.\ This phenomenon is commonly referred to as the Case\,C to D transition.\ Soft X-ray lines can be enhanced or suppressed up to $\sim$ 50 times as a consequence of this transition \cite{chakraborty22}.\ {\bf Figure~\ref{fig:caseABCD}} shows the contrast between Case\,C and Case\,D simulated using the recent version of \textsc{Cloudy} \cite{Chakraborty20a, Chatzikos23} for a power-law spectral energy distribution using the \lem\ responses for a 30\,ks exposure time.

\begin{figure}[ht!]
\centering
    \includegraphics[width=0.5\textwidth]{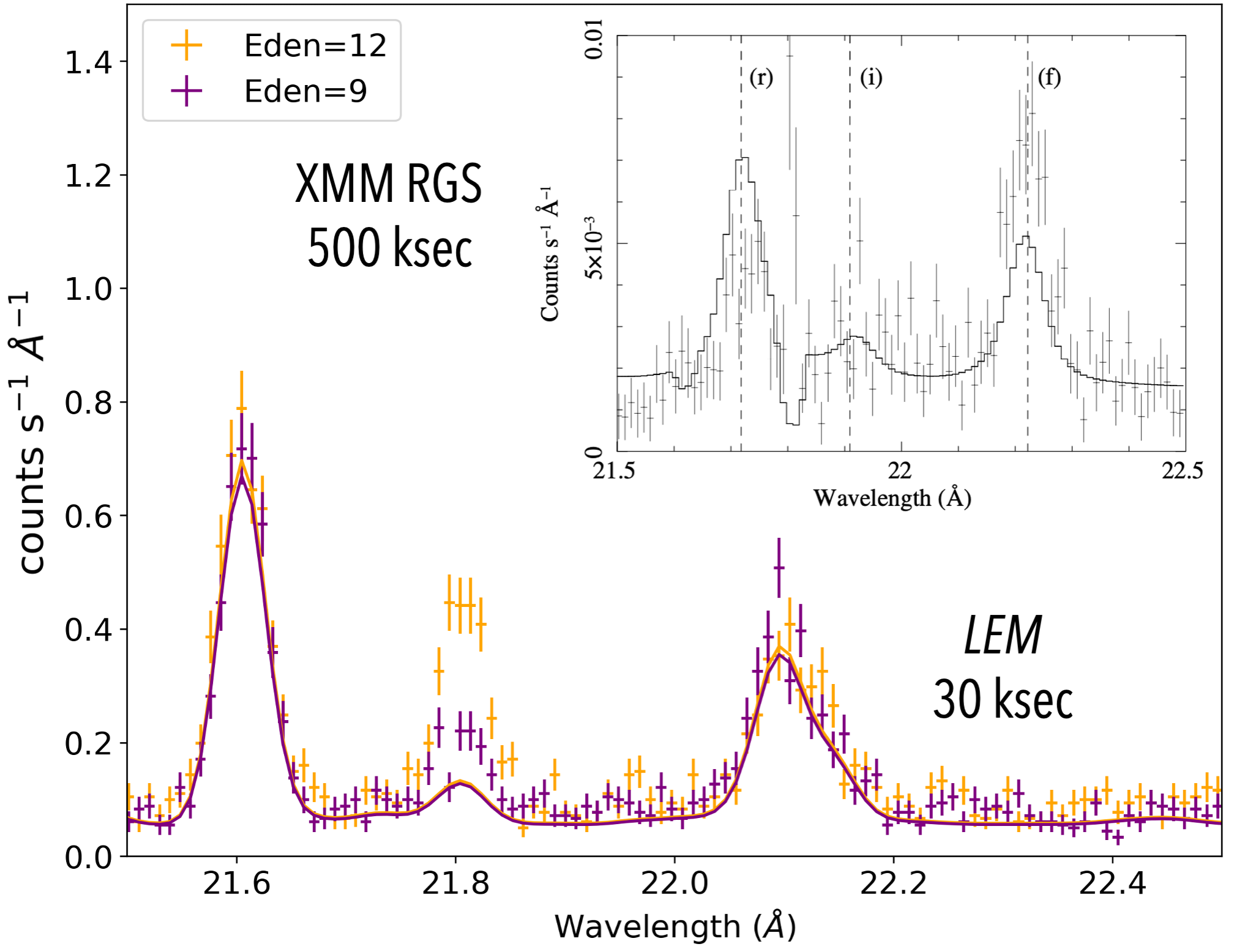}
    \caption{\footnotesize{{\bf \lem\ constrains photoionized plasma and distinguishes AGN from stellar activity in NGC 1365.}
     \lem\ simulation for 30\,ks of a thermal plus photoionized plasma for Case\,C and $U$=1, representing the implied conditions for the Seyfert\,2 and Starburst galaxy NGC\,1365\cite{guainazzi09}.\ Orange and violet points indicate an electron density log\,$n_e / {\rm cm}^{-3}=$ 12 and 9, respectively.\ The solid lines represent the collisionally-ionized component at kT=0.55\,keV, similar to the model in the Inset:\ O\,VII triplet obtained from a 500 ks \XMM/RGS observation fitted only with a collisionally-ionized plasma model.\ Significant residuals in the RGS indicate the presence of photoionized gas but n${_e}$ is not constrained.\  \lem\ constrains the photoionized plasma better than \XMM\ and with much shorter exposures - which gives us variability information on timescales of hours that the \XMM/RGS cannot provide.}}
    \label{fig:photo}
\end{figure}

The optical narrow-line region (NLR) with sizes of a few hundred pc to $\gtrsim$1 kpc and the optical broad-line region (BLR) on smaller scales are ideal locations to examine Case C/D and explore the connection between an AGN and its host galaxy.\
Optical narrow line diagnostics are historically used to discriminate between AGN and SF \cite{heckmanbest14}, and there is a great promise in using soft X-ray emission lines, which are typically found in Seyfert\,2 type AGN\cite{guainazzibianchi07, kinkhabwala02, bogdan17, gomezguijarro17} and can also be detected in Seyfert\,1s when the AGN enters a low flux state\cite{grupe08}.\ However, even deep exposures of well-studied nearby AGN with current instrumentation can only offer loose constraints on gas temperature, electron density (n${_e}$), and ionization parameter ($U$ = ionizing photon flux / $c$ n${_e}$, where $c$ is the speed of light).\ A prime example of the state-of-the art for such an observation is provided by a 500\,ks \XMM/RGS observation of NGC\,1365\citep{guainazzi09} ({\bf Figure~\ref{fig:photo}}).

\lem\ will place tight constraints on $U$ and $n_e$ in a fraction of the time compared to \XMM. Using {\sc cloudy} simulations we calculated a photoionized plasma model for
Case\,C\cite{chakraborty21} - see {\bf Figure~\ref{fig:photo}}.\ For the assumption of a spectrum similar to NGC\,1365\citep{guainazzi09}, including the thermal starburst component, we derived comparable errors to \XMM\ on $U$ in only 30\,ks (1/16 of current exposure times), and tight constraints on $n_e$.\ Defining the density diagnostic as $R$(n${_e}$) = $F$/$I$\cite{porquet10} (with $F$ and $I$ the fluxes of the forbidden and intercombination lines, respectively), we derive errors on $R$ from $1\%$ to $6\%$.\ This demonstrates \lem’s capability to constrain the conditions of photoionized plasma {\it on short timescales of hours or less}, even when mixed with a strong starburst.

The power of \lem\ is particularly notable in studies of low mass (e.g., $M_{\rm BH}$ $<$ ${10^5}$ M$_\odot$) and low luminosity (L$_{2-10 \, \rm keV} < 10^{41}$ erg s$^{-1}$) AGN that are missed in most surveys\cite{azadi2017}.\ In particular, the X-ray emission from an intermediate mass black hole is expected to be low and comparable to, and therefore indistinguishable from, stellar processes (e.g., high-mass X-ray binaries at $L_X\sim{10^{35} - 10^{40}}$ erg s$^{-1}$)\cite{ranalli2003}.\ This is particularly true in low metallicity dwarf galaxies, where the emission from stellar processes can be enhanced by over an order of magnitude\cite{fragos2013}.\ These galaxies represent pristine local analogs to the high redshift hosts of observationally inaccessible SMBH seeds and provide particularly strong insight into this population\cite{mezcua2019}.\ 

As {\sl eROSITA}’s all-sky survey at its full depth is expected to detect $\sim1,350$ X-ray sources in dwarf galaxies\cite{latimer2021}, and in the first data release has currently reported the detection of  $\approx 200$ in the full sky \citep{2024arXiv240601707S,bykov2024}, this ambiguity means that {\it eROSITA} alone is unable to robustly confirm the nature of these sources.\ Traditional multi-wavelength AGN identification diagnostics in other wavelengths similarly lose reliability in the low mass regime\cite{condon1991,trump2015,hainline2016,cann2019,satyapal2021}.\ Since their X-ray luminosities are indistinguishable from stellar sources, high-resolution spectroscopy is the only avenue to distinguish these targets using solely X-ray data, as there are predicted to be notable differences in both the continuum shape and emission line features of the spectra of these AGN compared to stellar sources.\ \lem\ will therefore play a significant role in confirming the nature of X-ray-emitting dwarf galaxies.\ We note that XRISM will provide an excellent opportunity to pave the way toward high-resolution spectroscopy but does not provide enough collecting area for this science.

\begin{figure}[h]
    \centering
    \includegraphics[width=0.5\textwidth]{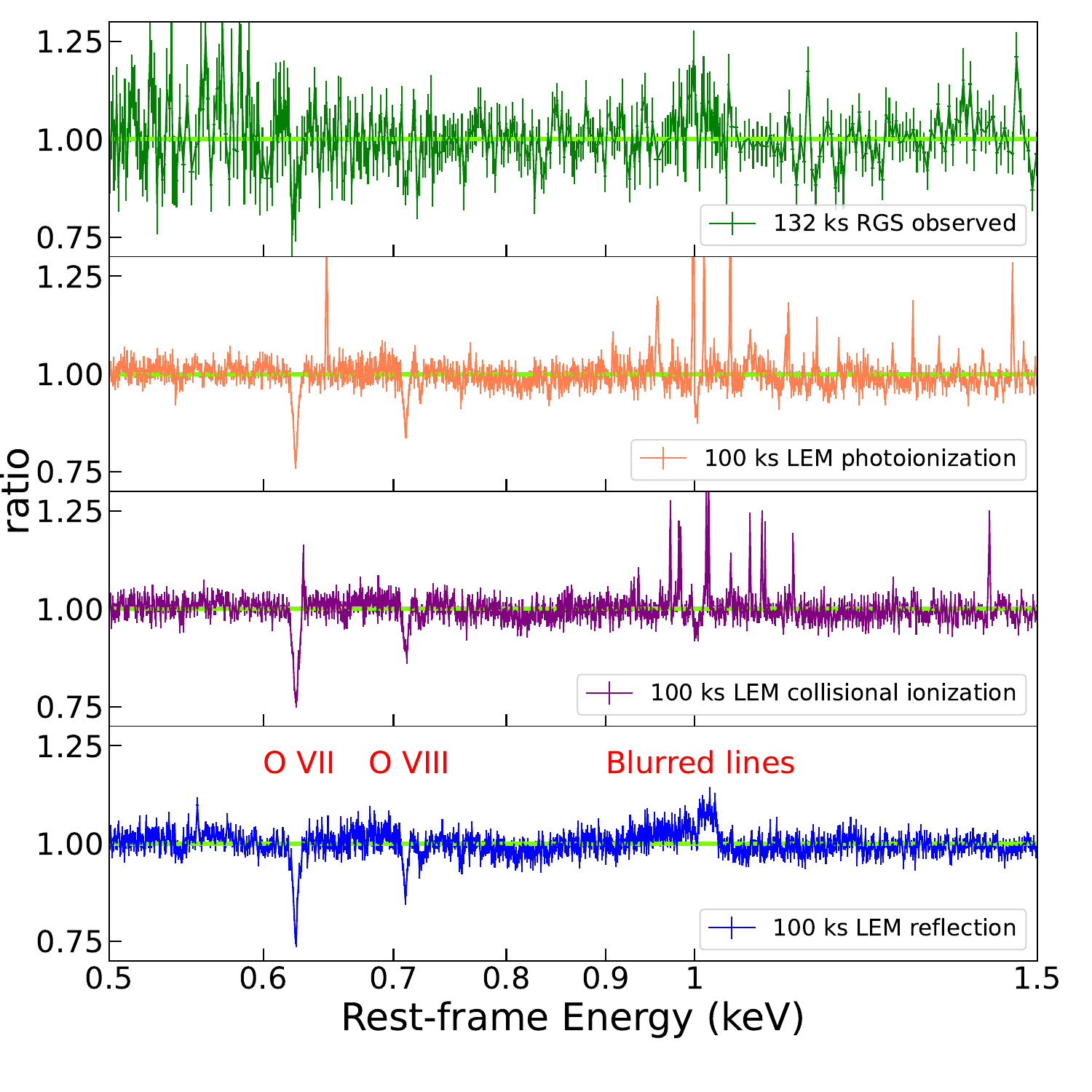}
    \caption{\footnotesize{{\bf Distinguishing different physical scenarios for SMBH accretion.} \LEM\ will be able to detect and resolve individual plasma lines, distinguishing different physical mechanisms. Simulations of three different cases - photoionization, collisional ionization, and reflection - are performed for 100~ks for the narrow line Seyfert\,1 1H\,1934-063 \cite{xu2022}.\ The \XMM/RGS data (top panel) cannot distinguish between these three cases.}}
    \label{fig:nlsy1a}
\end{figure}

\subsubsection{Black hole accretion and feedback on small scales.}
\label{sec:smallfeedback}

Seeking to determine how a SMBH is coupled to its host galaxy and to the nature of galaxy evolution leads us to ask two fundamental questions.\ What is the growth rate of the black hole and what is its impact in terms of outflow and kinetic rate onto the surrounding galaxy? \lem\ will unveil the disk-wind connection for highly-accreting SMBH and accretion-driven outflows, thus placing constraints on the outflow and net accretion rates.\ This in turn tells us how the SMBH affects its galaxy.

\begin{figure}[ht]
    \centering
    \includegraphics[width=0.5\textwidth]{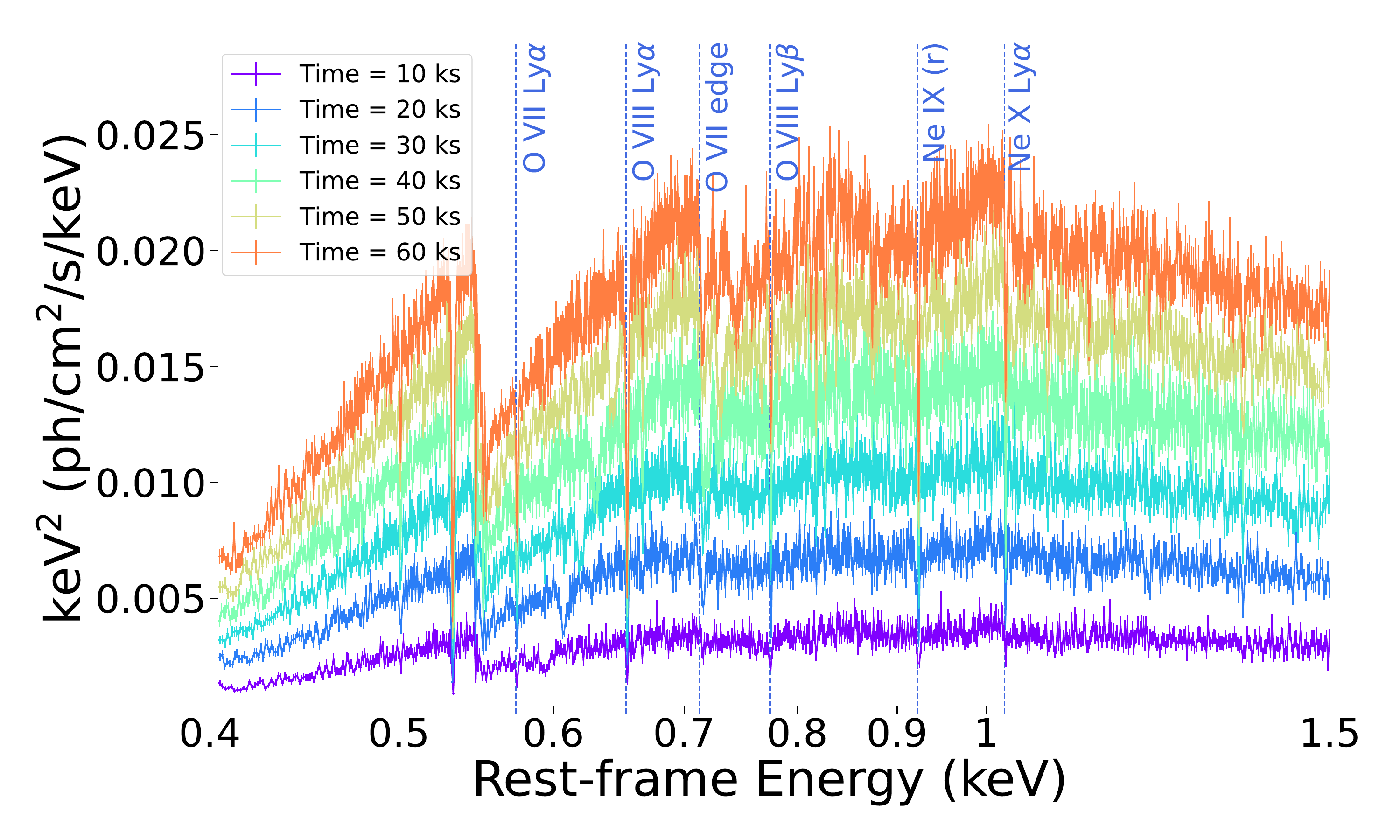}
    \caption{\footnotesize{{\bf \lem\ maps the response of accretion disk winds to SMBH luminosity.} \lem\ simulations of accretion disk wind response for a series of count rates for the narrow line Seyfert\,1 1H\,1934-063.\ {\bf In just 10\,ks, \lem\ quickly captures the response of the powerful, ionized AGN outflows to changes in black hole luminosity (accretion) via absorption lines.}\ Such short timescales are impossible to probe with current instruments but are critical to enable measurements of the currently elusive wind density and unravel the mysteries of wind launching mechanisms and the accretion disk-wind connection.\ The model consists of blueshifted photoionization in absorption and reflection in emission\cite{xu2022}.}}
    \label{fig:nlsy1b}
\end{figure} 

AGN winds are present in several forms with velocities ranging from $\sim100$ to $\sim$100,000\,km\,s$^{-1}$.\ These winds are important due to their effects on the accretion of matter onto and, therefore the growth of, the SMBH as well as the evolution of the surrounding interstellar medium.\ Currently, powerful winds are mainly studied via Fe\,XXV and Fe\,XXVI blueshifted absorption lines\cite{sim2008,matzeu2023}; however, a new form or phase of these winds is found via lower-ionization lines in the soft X-ray band, especially in highly accreting SMBH powering narrow-line Seyfert\,1 AGN\cite{hagino2016,Parker2017} and in some high accretion rate quasars such as PDS\,456\cite{reeves2018} ($M_{\rm BH}\sim{10^9} M_{\odot}$).\ These discoveries have opened a new channel to study the structure of AGN winds, such as the wind launching mechanism(s), the disk-wind connection, and the overall structure of the accretion disk.
\begin{figure*}[h!]
\centering
\includegraphics[width=\hsize]{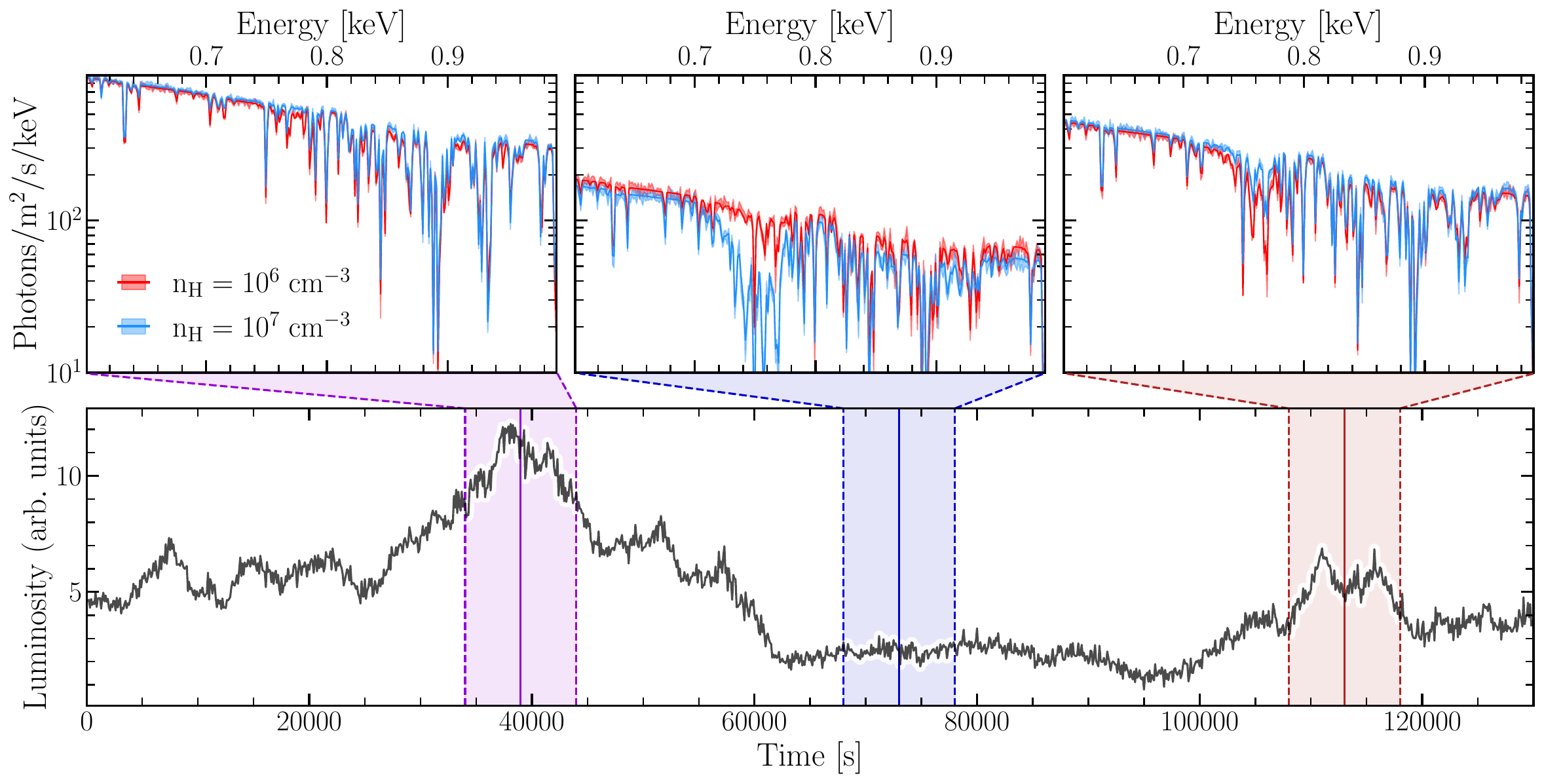}
\caption{\small {\bf \lem\ watches the spectral evolution of an absorbed AGN.} \lem\ simulation of the evolution of two typical AGN outflows with the same column density and ionization parameter but different density. We compared the spectrum at three different epochs along the changing X-ray light curve of a standard Seyfert\,1 galaxy. The synthetic spectra are generated using the time-dependent photoionization model \textsc{tpho} with an exposure time of 10~ks.}
  \label{fig:lem_simu}
\end{figure*}

Current instruments lack the necessary combination of effective area and spectral resolution to constrain the soft X-ray properties of disk winds within the continuum variability timescales, requiring stacking among different epochs\cite{Pinto2018}.\
\lem\ instead will detect and resolve individual plasma lines of the spectral features imprinted by the winds.\ This will test different explanations for the soft X-ray excess and emission around 1\,keV, ascribed to reflection off the wind base but unresolved with the {\it XMM-Newton}/RGS\cite{xu2022}.\ Blueshifted absorption lines (i.e. O\,VII--VIII) will be detected at $>5\,\sigma$ with a \lem\ exposure time of 10\,ks (see {\bf Figure~\ref{fig:nlsy1a}} and {\bf Figure~\ref{fig:nlsy1b}}) and will enable us to determine a) the nature of the ionized plasma, b) its ionization state and c) its response to the continuum changes within the variability timescales. 

The combination of \lem's sensitivity and spectral resolution will improve our understanding of the mechanism and launching regions of AGN winds and their mass outflow rates with high significance.\ \lem\ will also expand the number of viable targets beyond the $\sim10$ that can currently be studied at all with the {\it XMM-Newton}/RGS, likely accessing hundreds of fainter AGN and thus creating larger samples of nearby AGN for this science.

Spectral timing analysis can provide both precise estimates of the density and accurate spatial distributions of different ionized outflows in nearby AGN by probing the plasma response to changes in the ionizing radiation \cite{Nicastro1999,Kaastra2012}.\ Fluctuations in the ionizing luminosity can cause a departure of the photoionization equilibrium in the surrounding plasma.\ The recombination timescale ($t_{\rm rec}$) quantifies the time necessary for a medium to reach the equilibrium after a variation in ionizing radiation, and it critically depends on the characteristic density of the plasma ($t_{\rm rec} \propto n^{-1}$).\ Consequently, low-density gases are more likely to be found out of equilibrium, especially if the variability timescale of the ionizing source is shorter than the recombination timescale of the photoionized plasma. 

\begin{figure*}[h!]
\centering
\includegraphics[width=\hsize]{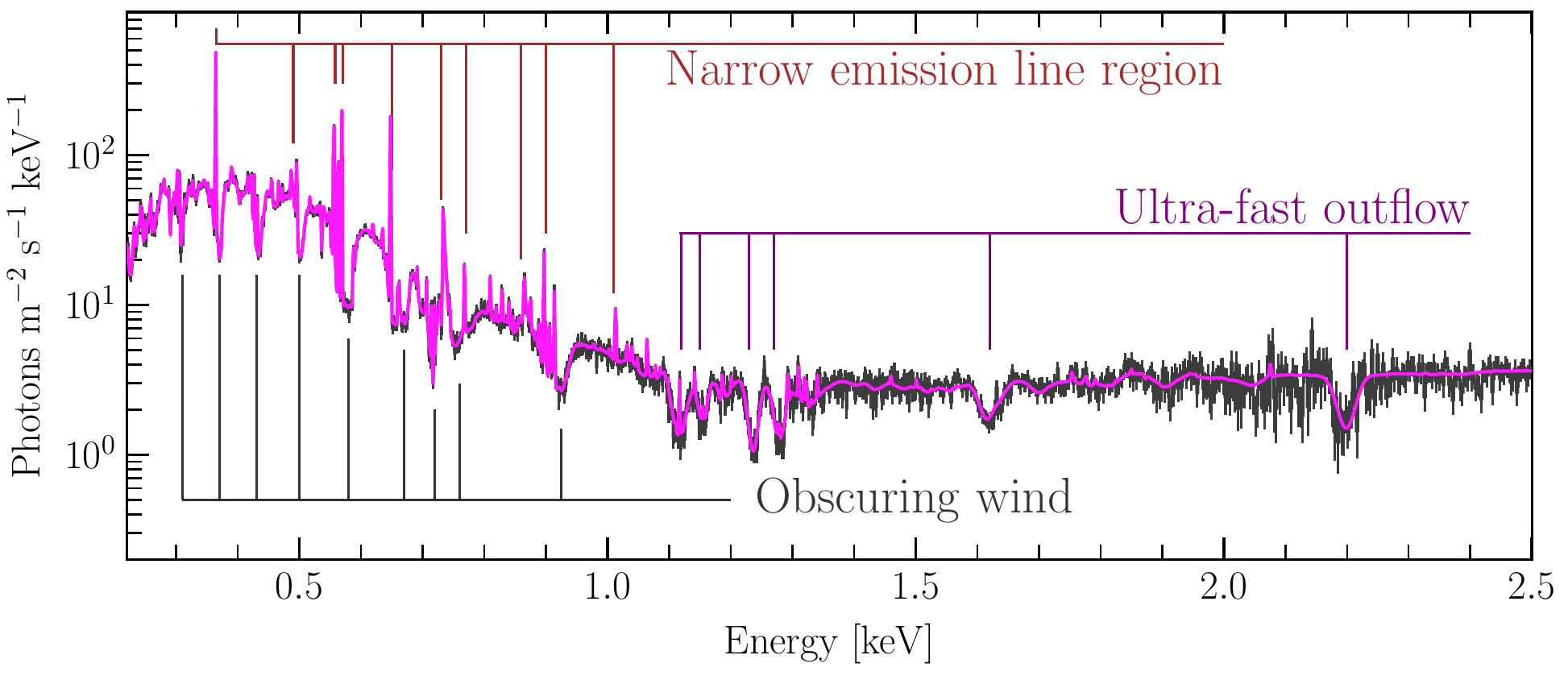} \\
\includegraphics[width=\hsize]{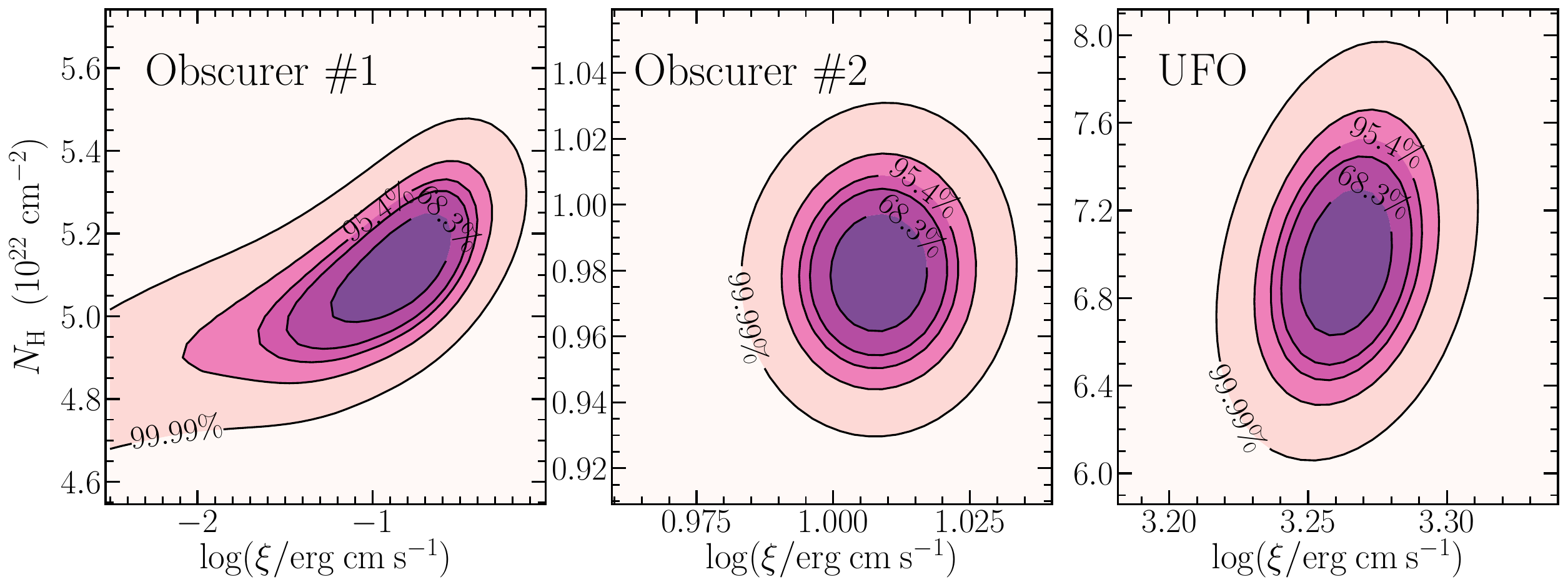} \\
\caption{\small {\bf \lem\ places tight constraints on outflow properties.} {\it Top panel}: \lem\ simulation of a typical multiphase obscurer outflow observed in multiple Seyfert I galaxies. We assumed an exposure time of 100~ks. The \lem\ synthetic spectrum is binned for clarity. {\it Bottom panel}: LEM will constrain the ionization parameter, $\xi$ (in log units), and the column density of each obscurer component. In the inner-outer direction, we plot the 68.3\%, 90\%, 95.4\%, 99\%, 99.99\% confidence constraints on the key parameters.}
  \label{fig:lem_obs}
\end{figure*}

{\bf Figure~\ref{fig:lem_simu}} compares the spectral evolution of an AGN absorbed by outflows with two different densities calculated with the time-dependent photoionization model \textsc{tpho} \cite{Rogantini22b}.\ The distinct evolutions of the spectra confirm the capability of \lem\ together with the new time-dependent photoionization model to determine the plasma density.\ The large collecting area of the microcalorimeter will allow us to probe the response time of high-density gases of $10^8$ to $10^{11}$ cm$^{-3}$, which vary on shorter timescales, likely located closer to the ionizing source.\ Precise estimates of the density and the distance provide tight constraints on the energetics of AGN outflows and therefore on their impact on the evolution of the host galaxy via AGN feedback\cite{2012ARA&A..50..455F}.\ Moreover, by comparing the kinetic power and momentum it will be possible to unveil the connection between different X-ray outflows such as the fast (sub-relativistic) highly ionized winds, obscuring clouds, and slow ionized outflows (historically known as warm absorbers). 

As shown recently for the bright AGN Mrk\,110\cite{reeves2021}, the detection of variable soft X-ray O\,VII emission lines, both in profile and intensity on timescales down to two days, using {\sl XMM-Newton}/RGS spectra provides us a complementary tool (compared to the Fe\,K$\alpha$ line at 6.4\,keV in the AGN rest-frame energy) to determine the inner radius of the relativistic reflection emission, ionization state, and inclination of the accretion disk.\

Several bright, local, Seyfert I galaxies have shown in the last decades to have strong obscurer outflows (e.g. \cite{Kaastra2014,Kara2021}).\ This obscuring wind significantly suppresses the X-ray flux and imprints strong, broad, and blueshifted absorption lines, such as C \textsc{iv} in the UV spectra taken with HST.\ The HST/COS data suggest that these obscurers outflow with a velocity of several 100s to 1000s of km s$^{-1}$, are located in the proximity of the BLR, and shield the surrounding medium, (\cite[e.g. warm absorbers][]{Mehdipour24}).\ So far, they appear as transient events with durations over timescales from days \citep{Risaliti11, Wang22} to several years \citep{Mehdipour22}.\ In several cases, the obscuring wind consists of multiple components \citep{Kaastra2014, Kriss19}.\ However, their properties, such as the ionization states, column density, and velocities, are poorly understood from X-ray absorption lines due to either poor signal-to-noise ratio or resolution. LEM will detect and resolve multiple absorption lines imprinted by the obscuring winds.\ {\bf Figure \ref{fig:lem_obs}} shows the 100~ks LEM simulation of a two-component obscurer in a bright Seyfert I galaxy with typical parameter values: column density $N_{\rm H} = 1-5\times 10^{22}\rm\; cm^{-2}$, ionization parameters $\log \xi$ between -1 and 1, outflow velocity $v_{\rm out} = 2000-3000$ km s$^{-1}$, turbulence velocity $v_{\rm turb} = 2000-3000$ km s$^{-1}$, and covering factor $f_{\rm cov} = 0.7-0.9$. We also added a two-ionization zone narrow emission line region with $\log \xi = 1-2.5$, which usually arises when the primary continuum is heavily absorbed \cite{Parker19}.\ LEM will also constrain the properties of a potential UFO through the detection of high-ionization absorption lines such as O~\textsc{viii}, Mg~\textsc{xii}, Si~\textsc{xiv} and S~\textsc{xvi}.\ These spectral features are evident in the 100~ks LEM synthetic spectrum ({\bf Figure~\ref{fig:lem_obs}}) where we included a UFO with $N_{\rm H} = 7\times 10^{22}\rm\; cm^{-2}$, $\log \xi=3.3$, and $v_{\rm out} = 0.1c$. 

The information that LEM will provide regarding the properties of warm absorbers, obscuring winds and ultra-fast outflows, together with what we have learned from UV spectroscopy, will give a comprehensive understanding of AGN ionized outflows. The synergy between UV, soft and hard X-ray observations (with, respectively, HST/COS, LEM and XRISM) will answer the long-standing questions on the origin of these different AGN outflow components and their relation to each other.

\begin{figure*}[ht!]
 \centering
\includegraphics[width=\hsize]{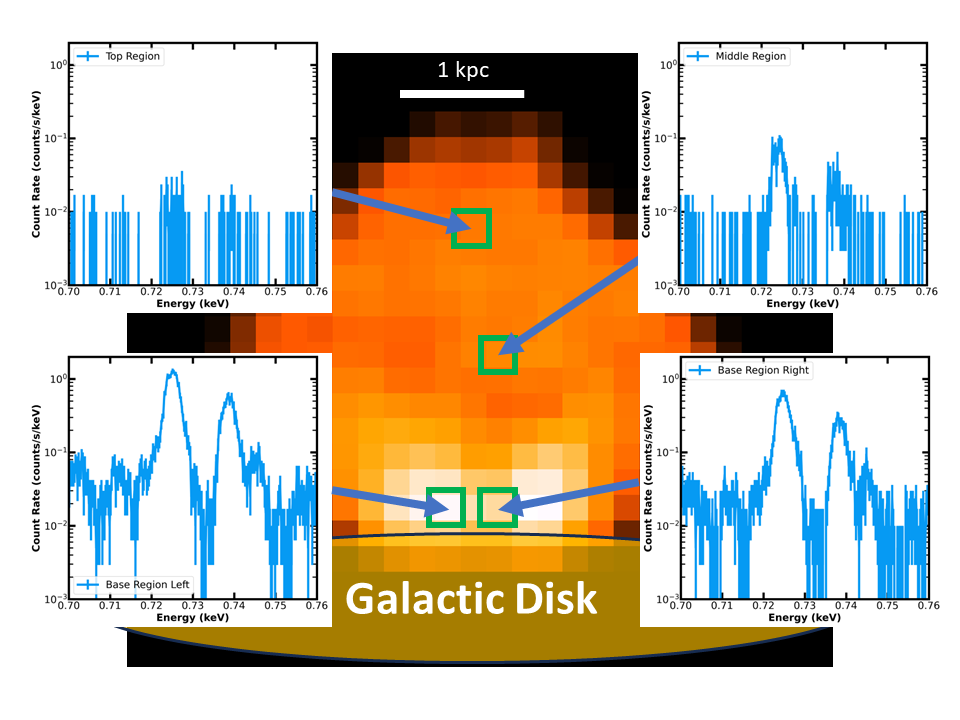}
\caption{\small {\bf Spatially mapping AGN-driven and starburst-driven outflows.} \lem\ simulations of spatially resolved spectra for a nearby (z = 0.001) starburst outflow with an AGN component on kpc scales with an exposure time of 100 ks.\ The simulated image shows the top half of a bi-conical galactic outflow.\ The non- X-ray emitting disk of the $L_{*}$ galaxy is at the base of the outflow.\ The spectra show Fe\,XVII lines extracted from four different pixels as indicated by the green squares. The X-ray emission from the disk of the galaxy has been masked out and this figure shows purely the emission from the outflowing gas. Velocities range from $\sim1,000$ km s$^{-1}$ near the disk to $\sim100$ km s$^{-1}$ at the edge of the bubble.}
  \label{fig:lem_outflow}
\end{figure*}

\subsubsection{Large Scale Outflows and feedback.}
\label{largefeedback}

In galaxies, kpc-scale, multi-phase outflows are often observed.\ These outflows can be generated by episodes of star formation \cite{Rupke18} or by AGN \cite{Veilleux2005}.\ Knowing the mechanisms that produce these outflows/winds is crucial to understanding the processes of galactic feedback and the enrichment of galactic halos, which forms a primary science pillar for \lem.\ Starburst- and AGN-generated winds differ due to their temperature and the amount of mass loading into the wind \cite{heckman2023}, and X-ray observations are well-suited to accurately detect and measure the hot gas in these outflows.\ Pure starburst-driven outflows have been well studied in the X-rays \cite{heckman2023}, but less work has been done on outflows with an AGN-driven component.\ With its combination of effective area, energy resolution, and grasp at low energies, \lem\ will make significant advances in probing the physical properties of these outflows and seeking to measure the observational signatures in the presence of an AGN.\ 

In running simulations including an AGN\cite{tanner2024}, we find that the biggest difference between a pure starburst outflow and a pure AGN outflow is the fraction of cold gas in the wind.\ A pure AGN-driven outflow has little cold gas embedded in it, and thus little soft X-ray emission, which would come mainly from shock heating by the wind on gas clouds.\ An AGN outflow with no corresponding star formation is almost invisible in soft X-rays since about $\sim$97\% of the outflow mass is made up of the hottest gas. On the other hand, a composite AGN-starburst outflow has the potential to provide key diagnostic signatures in the \lem\ bandpass.

For very nearby galaxies, \LEM\ can directly measure spatial variations such as velocity structure across an outflow ({\bf Fig.~\ref{fig:lem_outflow}}).\ This kind of dynamical information is impossible with CCD spectroscopy or dispersive gratings because they cannot deal with dim, spatially resolved sources.\ To create simulated \lem\ observations from extended emission we used a hydrodynamic simulation of an AGN wind-driven galactic outflow ($P_{AGN} = 10^{42}$ erg s$^{-1}$) combined with nuclear star formation ($M_{*} = 10^6 M_{sun}$) for an $L_{*}$ galaxy\cite{tanner2024}.\ Here, $P_{AGN}$ is the total power of the AGN which includes the kinetic energy of the wind and radiation.\ We assume a distance of $z = 0.001$ to the nearby galaxy.\ The initial gas distribution in the galaxy model assumes a two-phase medium with a clumpy, cold, dense galactic disk embedded in a hot halo.\ 

The combined AGN wind and nuclear starburst create a three-phase outflow characterized by a very hot volume filling component.\ The outflow consists of cold, dense gas clouds pushed out of the galaxy disk by ram pressure, warm gas ablated from the cold clouds, and hot gas.\ The hot gas can be further broken down into cool, warm, and hot components.\ The coolest hot gas phase produces X-ray emission and comes from gas being ablated off of the dense clouds, or gas swept up in shocks. The warm, hot gas is the hottest of the shocked gas, and gas in the later stages of mixing with the hottest part of the outflow. The very hot, hot gas fills the majority of the outflow volume.\ X-rays arise from gas in the bubble along with shock-heated gas at the boundaries of the bubble and the transition region between the hot wind and the dense clouds.

X-ray images and spectra were generated using a combination of py{\sc XSIM}\footnote{\url{https://hea-www.cfa.harvard.edu/~jzuhone/pyxsim/}} and {\sc SOXS}\footnote{\url{https://hea-www.cfa.harvard.edu/soxs/}}.\ In {\bf Figure \ref{fig:lem_outflow}}, we show a full band image of the resulting outflow assuming an exposure time of 100\,ks.\ The insets show the distinct spectra from individual \lem\ pixels - dominated by the starburst outflow - and we have plotted Fe\,XVII lines from each region.\ Near the base of the outflow the two selected regions have similar line profiles with different intensities.\ There are small differences in the line shapes due to the complex kinematic structure of the gas being modeled.\ Moving up through the outflow, the line intensity decreases and velocities decrease.\
In this particular example, the diminishing line profile indicates a decreasing temperature gradient through the bubble.\ The temperature profile of the outflow can be used as a diagnostic of different radial cooling profiles of the wind.\ Only with the resolution of \lem\ can individual lines be separated in this way over a wide observational field to provide this type of diagnostic.

\begin{figure*}[h]
\centering
\includegraphics[width=\hsize]{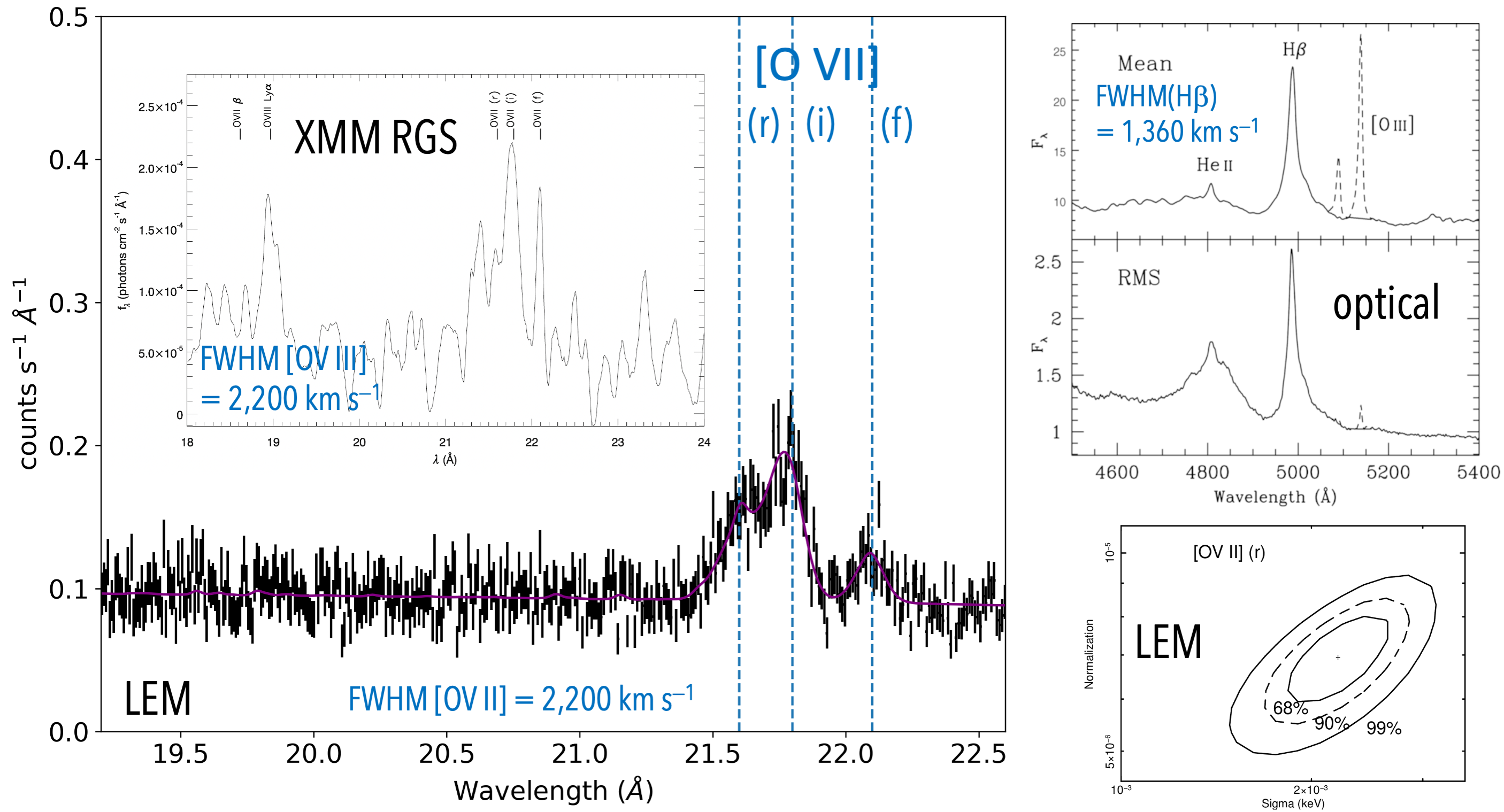}
\caption{\small {\bf Optical and UV techniques can be applied with \lem\ to perform reverberation mapping for the BLR.} Left: This figure illustrates a \lem\ 60\,ks simulation for Mrk\,335 assuming O\,VII from the BLR with FWHM = 2,200 km s$^{-1}$.\ Inset shows the {\sl XMM-Newton}/RGS spectrum\cite{longinotti2008}, which has a measured O\,VIII Ly$\alpha$ FWHM = 2,200 km s$^{-1}$.\ Right Top: Mrk\,335 optical mean (top) and rms (bottom) spectra; figure reproduced from \cite{Grier2012}.\ The optical spectra are flux-calibrated in the observed frame ($z$ = 0.0258) with flux density in units of $10^{-15}$ erg cm$^{-2}$ s$^{-1}$ Angstrom$^{-1}$. Bottom right: \lem\ confidence contours for our simulated recombination line for line width ($\sigma$) vs.\ normalization.} 
  \label{fig:reverb}
\end{figure*}

\subsection{Soft X-ray studies comparable to optical/UV bands: BLR, NLR kinematics, reverberation mapping}
\label{sec:reverb}

With its $\sim$100 km s$^{-1}$ energy resolution and large collecting area, \lem\ compares well with optical and UV spectroscopy and will measure properties of individual soft X-ray emission lines that may arise from the broad line region (BLR) or narrow line region (NLR).\ \lem\ will not only resolve these lines and provide plasma diagnostics (Section \ref{sec:photo}) and kinematic information, but will also map their variability on less than 100\,ks timescales, a significant improvement upon the {\sl XMM-Newton}/RGS.\ For the first time, this variability can be directly compared with optical/UV spectra to examine relative variability properties between these wavebands.\ Soft X-ray reverberation mapping - measuring the response of the emission lines variations in the continuum flux - can be performed for individual narrow and broad emission lines, providing insights on the structure of multiple emitting regions.

For emission lines dominated by viral motion, time delays in their response to the continuum can determine the sizes of the emitting regions while time delays combined with line widths provide a viral mass for the SMBH \cite{peterson2004}.\ To illustrate this science with \lem\, we have chosen the NLSy1 Mrk\, 335 ({\bf Figure~\ref{fig:reverb}}) which has soft X-ray lines consistent with a photoionized gas.\ In a very low flux state, a 25~ks observation with the {\sl XMM-Newton}/RGS reveals strong, soft X-ray emission lines not detected in previous, higher flux states, such as highly ionized Fe lines, O\,VII, Ne\,IX and Mg\,XI lines\cite{longinotti2008}.\ The most prominent group is the complex of O\,VII emission lines around 22\,\AA.\ As discussed in Section \ref{sec:photo}, line ratios in He-like ions directly probe properties of the emitting gas.\ 

The O VII triplet is resolved with the {\sl XMM-Newton}/RGS although individual line widths are not constrained ({\bf Figure~\ref{fig:reverb}}, inset).\ {\sl XMM-Newton} provides $\sim50-60$\% errors on the diagnostic line ratios\cite{longinotti2008}, but with the line energies and widths held fixed.\
The most well constrained feature is the O VIII Ly$\alpha$ line with FWHM = $2,200 \pm 750$ km sec$^{-1}$,
providing a distance estimate of $\sim 2.3\times10^{16}$ cm to $1.2\times10^{17}$ cm, consistent with photoionized gas from the BLR\cite{longinotti2008}.\ The optical spectrum of Mrk\, 335 shows FWHM(H$\beta$ $\lambda4861$) = 1,360 km s$^{-1}$ and FWHM(He II $\lambda4686$) = 3,200 km s$^{-1}$ \cite{Grier2012} with line-emitting regions at distances of $3.6\times10^{16}$ cm and $7\times10^{15}$ cm for H$\beta$ and He II, respectively.\ The time lags for H$\beta$ and He II relative to the optical continuum are $13.9\pm0.9$ and $2.7\pm0.6$ days, respectively. Combined with the optical line widths, these lags provide a black hole mass estimate of $3\times10^7$ solar masses\cite{Grier2012}. 

Using the best-fit model\cite{longinotti2008}, from the {\sl XMM-Newton}/RGS we simulated a \lem\ spectrum for 60~ks
({\bf Figure~\ref{fig:reverb}}).\  We replicated the {\sl XMM-Newton} spectrum with a power law ($\Gamma$ = 2.7) and individual Gaussian emission lines for the O VII triplet. We additionally  set the O VII line widths to 2,200 km s$^{-1}$ (the measured FWHM of O VIII Ly$\alpha$) and assigned the relative line fluxes from the RGS. \lem\ constrains energy, width, and intensity for all three features. 

We discuss the recombination (r) and intercombination (i) lines, blended at \lem’s resolution. We recover FWHM(O VII, r) = $2,640\pm320$ km s$^{-1}$ (12\% errors at 90\% confidence) and FWHM(O VII, i) = $2000\pm132$ km s$^{-1}$ (7\% errors at 90\% confidence).\ Plasma diagnostic line ratios assuming these velocity-broadened lines provide $\sim20-40$\% errors, a factor of $\geq2$ times better than {\sl XMM-Newton} for just 2.4 times the observing time (and in this case the line energies and widths can be measured).\ Line blending creates systematic errors that dominate the statistical errors on width and normalization, but this shows that proper line deconvolution techniques will be required - similar to optical spectral modeling\cite{mullanyward2008,tremou2015}.

\begin{figure}[t]
    \includegraphics[width=0.48\textwidth]{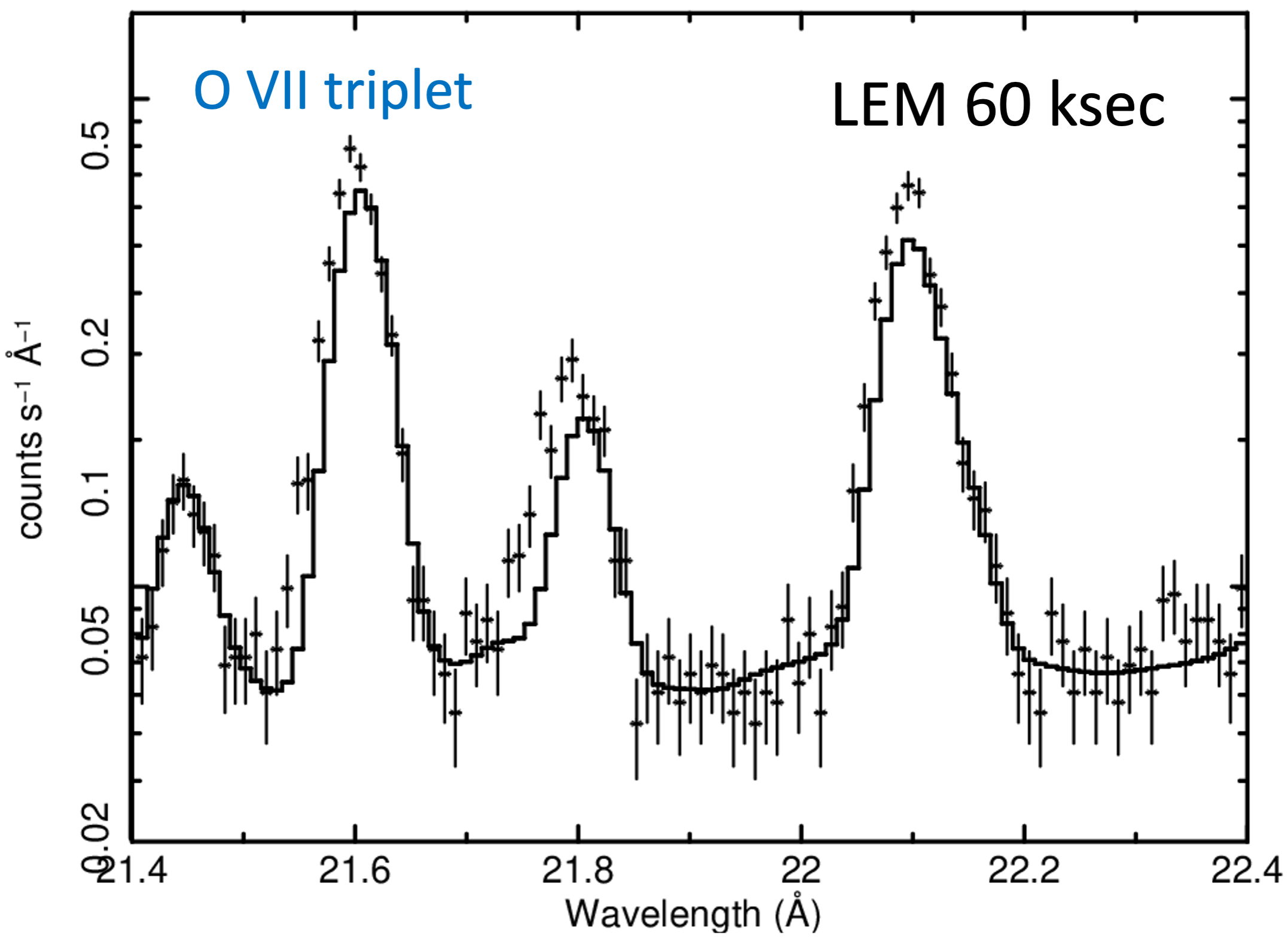}
    \includegraphics[width=0.48\textwidth]{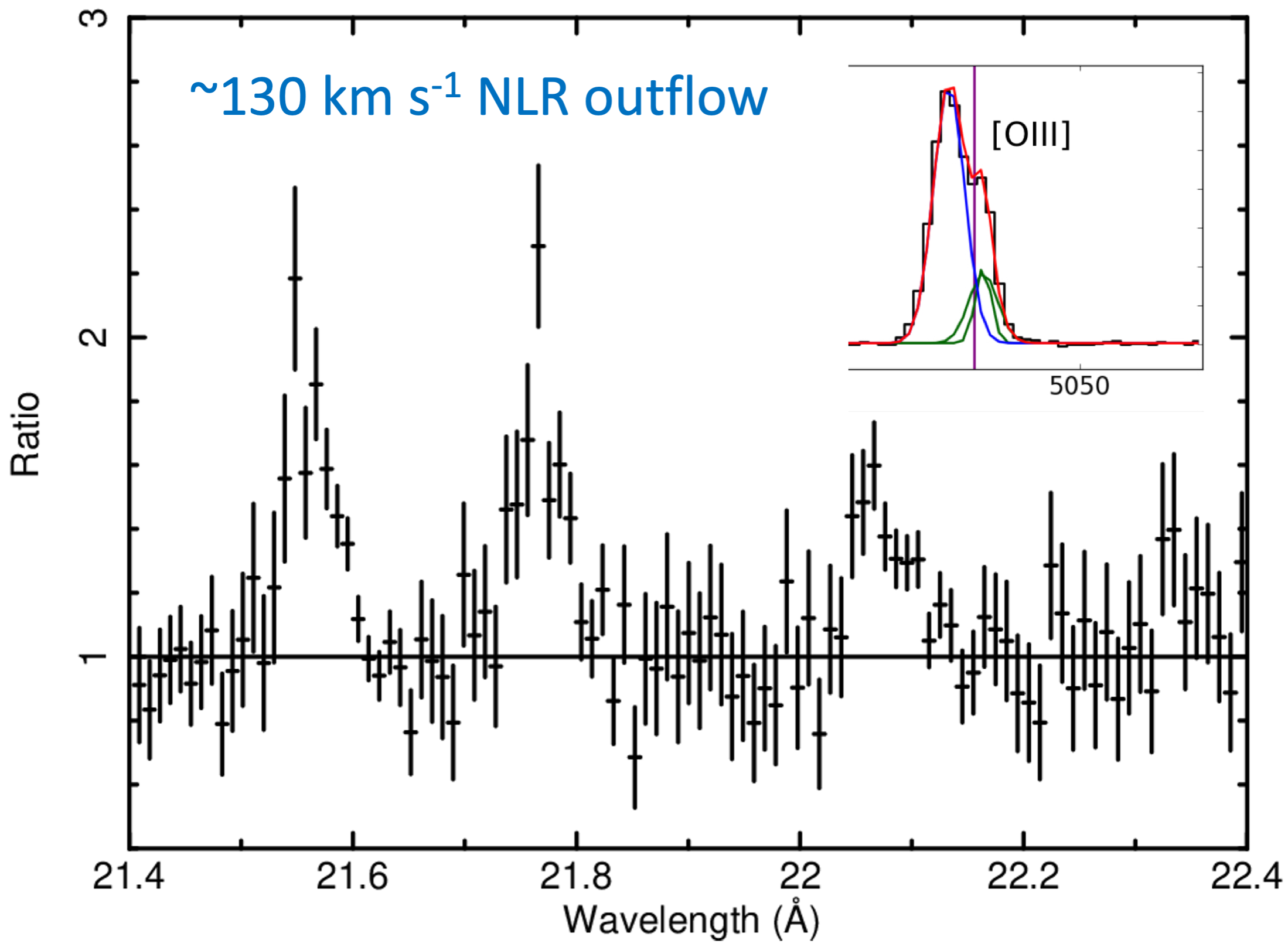}
    \caption{\footnotesize{{\bf \lem\ measures outflows in the NLR.}} At LEM's 1.3 eV energy resolution, information can be derived from the narrow line region (optical velocities of $\leq1,000$ km$^{-1}$). Top panel is a simulation of NGC 1365\cite{guainazzi09} (Section \ref{sec:photo}, Figure \ref{fig:photo}), this time adding emission from an outflow with v = 130 km s$^{-1}$ to match velocities in the optical NLR\cite{venturi2018}. Bottom: Ratio of the simulated spectrum to the model that only includes thermal and photoionized emission at the systemic velocity, leaving the signature of a blueshifted outflow feature. Inset is an optical VLT MUSE spectrum showing such a blueshifted feature\cite{venturi2018}. } 
    \label{fig:energy_offset}
\end{figure}

The studies possible with \lem\ can provide further insight into optical reverberation mapping programs.\ Current programs with ground-\cite{grier2017,guo2022,u2022} and space-based\cite{homayouni2022} observatories have provided precise mass measurements of a sample of $\sim$70 supermassive black holes to date\cite{bentz2015}. This sample is expected to increase dramatically in the onset of the {\sl Vera Rubin Observatory} and the 10-year {\sl Legacy Survey of Space and Time} (LSST), where over a thousand continuum reverberation mapping measurements are expected to become available\cite{kovacevic2022}.\ The measurements provided by \lem\ would likely probe different spatial scales than those accessible by current observatories, and open up new parameter space for studying multiple timescales.

Studies of the statistical properties of the long term soft X-ray variability of local AGN on the timescale of many decades can be performed by combining \lem\ observations with previous X-ray observations ({\sl ROSAT}, \XMM, {\sl Chandra}, {\sl eROSITA}).\ Variability can be investigated using the structure function analysis which describes the amount of variability as a function of the lag between the X-ray observations.\ The AGN structure function built using \ROSAT\ and \XMM\ data of 220 sources increases over $\sim$20 years at long time lags \cite{middei2017}. This suggests that variability in the soft X-rays can be influenced by flux variations originating in the accretion disk or that these variations take place in a region large enough to justify variation on such long timescales.\ With \lem, we will push these studies to longer timescales and search for evidence of a plateau in the structure function on the longest timescales, which has the potential of e.g. constraining the size of the soft X-ray emitting region in individual AGN.

\LEM\ will also complement studies of the NLR in AGN by detecting outflows with velocities of a few hundred km s$^{-1}$, even in the presence of thermal emission and/or photoionized gas at the systemic velocity. Optical outflows of up to 150 km s$^{-1}$ are seen with the VLT Multi Unit Spectroscopic Explorer (MUSE) in NGC\, 1365\cite{venturi2018}. We, therefore, repeated our {\sc xspec} simulation of NGC 1365 in Section \ref{sec:photo} with a multi-component APEC and CLOUDY model, this time with $U$ = 0 and n${_e}$ = 10, within conditions implied from the \XMM/RGS. We maintained the individual line fluxes and added blueshifted emission as a Gaussian corresponding to 130 km s$^{-1}$. The 60~ks \lem\ simulation is shown in {\bf Figure \ref{fig:energy_offset}}. Dividing out the emission lines at the systemic velocity, the ratio of the data to the model shows the blueshifted residuals and \lem's ability to easily detect energy shifts of $\ge130$ km s$^{-1}$ for X-ray bright AGN.

\subsection{AGN science at high redshift (epoch of re-ionization): Survey and GO Science}
\label{sec:agnhighz}

\begin{figure}[t]
    \includegraphics[width=0.48\textwidth]{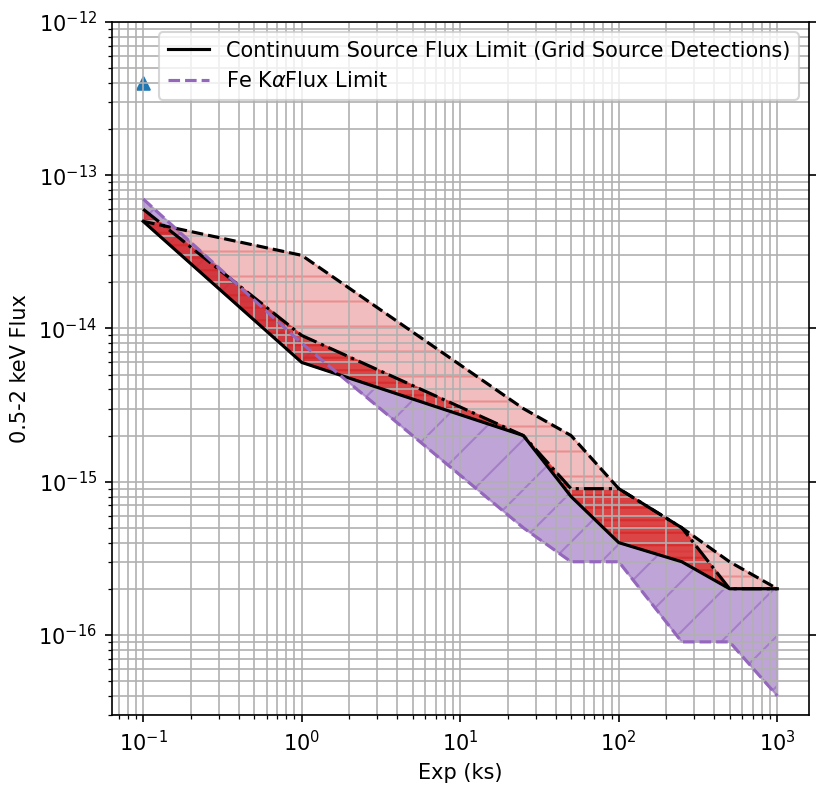}
    \caption{\footnotesize{{\bf Flux Sensitivity curves for \lem\ as a function of exposure time.} The solid black curve indicates the estimates for the full 0.5--2\,keV \lem\ energy band, the dashed black curve indicates the hard 1--2\,keV band, dash-dotted black line indicates the soft 0.5--1\,keV.\  Purple dashed line: flux limit for Fe\,K$\alpha$ emission lines observed in 100\,eV spectral windows.\ These flux limits are all derived via {\sc wavdetect} detection of sources with known 0.5--2\,keV fluxes.\ We shade the area between each set of curves to emphasize the range of limiting fluxes possible depending upon the user choice: solid red shading indicates the range between the 0.5--2\,keV and 0.5--1\,keV flux limits; light red shading indicates the range between the 0.5--2\,keV and 1--2\,keV flux limits; purple shading indicates the range between the 0.5--2\,keV and the Fe\,K$\alpha$ flux limits. }} 
    \label{fig:fluxsensitivity}
\end{figure}

Beyond the science topics discussed above, \lem\ will also address many unanswered questions at higher redshifts.\ Foremost are AGN population studies at $z\ge2.5$.\ The fraction of obscured vs.\ unobscured AGN can be determined from detections and comparisons of the Fe K$\alpha$ line strengths.\ Also, the AGN-starburst connection and how AGN are related to star formation rates and other galaxy properties can be examined from their collisionally-ionized features where we detect them (ionized Fe from hot gas) as well as using large multiwavelength samples to compare with optical, IR and UV properties.

The \lem\ All-Sky Survey of 100\,s depth (16\,Ms total) over the mission lifetime\citep{Khabibullin2023} and pointed deep fields (250-900\,ks on average across $\sim 16-20$\,deg$^2$) will provide rich opportunities for AGN science within the directed mission (Section \ref{subsec:src_detections}).\ Moreover, the GO program will offer additional opportunities for surveys and targeted observations of interesting local- and higher-redshift sources.

With unprecedented in-depth high-$z$ studies with the \JWST\ and the {\it Roman Space Telescope}, it will be vital that a selected X-ray probe can contribute to this ongoing conversation.\ \lem\ possesses unique capabilities to complement multi-wavelength efforts in this exploration of the epoch of re-ionization.\ Using {\it JWST}, we are able to observe the furthest reaches of the Universe possible so far, including galaxies up to redshift of $\sim$12\cite{finkelstein2022} and a detailed study of AGN hosts from $z \sim$3--5\cite{kocevski2023,Yang2023}.\ \lem\ is perfectly suited to provide complementary observations and high-energy X-ray insight into this population of host galaxies, allowing for more detailed study of the accretion properties of the AGN that reside inside them. In particular, while current progress in understanding these higher redshift samples can begin now with eROSITA, LEM will usher in both higher spatial resolution and higher spectral resolution coverage of these AGN populations with its comparable -- if not better -- effective area at restframe 1\,keV. These advancements in spectral and spatial resolution will reduce source confusion and allow the spectral disentangling of the complex emitting regions of higher redshift AGN populations because the hard power law continua and Fe K$\alpha$ emission lines of these AGN at $z>2.5$ will redshift down into the LEM passband.

Furthermore, a recent study from the {\it Cosmic Evolution Early Release Science} (CEERS) \JWST\ collaboration uncovered a significant number of $z>3$ obscured AGN host galaxies \cite{Yang2023} missed by current high-$z$ X-ray studies.\ While obscuration estimates for these targets have been reported, high uncertainty is acknowledged, as rest-frame hard X-ray observations are needed to precisely constrain the measurements. The redshift range of these targets is perfectly suited to a \lem\ study; the rest-frame hard X-rays necessary to detect these targets and measure obscuration are redshifted into the \lem\ energy band, allowing for a robust X-ray -- mid-infrared (MIR) study of these unique targets at a period of peak growth. 

There are also significant efforts to increase the populations of detected high-$z$ galaxies and AGN with current and future observatories for use in statistically large studies.\ For example, {\sl COSMOS-Web}\cite{casey2022} is expected to unearth nearly 1,000 galaxies with 6.5$<z<$7.5 with \JWST\cite{Marshall2022}.\ These studies will be further advanced with the advent of the {\sl Roman Space Telescope} and {\sl Euclid}, which will offer high sensitivity over wider fields of view, where yields of hundreds and potentially thousands of quasars at $z\sim$6--8 are expected to be discovered throughout their survey periods\cite{euclid19,marshall2020}.\ These targets are all ideal for \lem\ follow-up or potential inclusion into \lem’s deep field observations (see point source detectability in Section~\ref{subsec:src_detections}) to further explore the relationship between IR and X-ray properties in the early universe and in periods of rapid growth. 

\subsection{Individual source detectability}
\label{subsec:src_detections}

By the nature of its design, \lem\ is optimized for the detection of AGN of moderate-to-high luminosity to medium to high redshift in every field. Relatively few\cite{Vito2014,Kalfountzou2014,Georgakakis2015,Vito2018} X-ray AGN are known to redshifts beyond $z\gtrsim4-5$, making it difficult to place robust constraints on the number of X-ray AGN above a given flux level per unit area on the sky (known as the log[N]--log[S] relationship, \cite{Kalfountzou2014,Vito2018}) and co-moving volume estimates for X-ray AGN out to larger cosmic distances.\ Only about thirty AGN from the {\sl C-COSMOS} X-ray catalog \cite{elvis2009,civano2012} and \textit{Chandra} Multiwavelength Project ({\sl ChaMP}) X-ray catalog\citep{kim2007,green2009} were used\cite{Kalfountzou2014} about a decade ago to place constraints on log(N)--log(S) for $z>4$ AGN. More recent works have made improvements using a sample of $\sim100$ X-ray AGN with photometric redshifts to extend the log(N)-log(S) relation out to $3<z<6$ and down to fainter flux levels (7.8$\times$10$^{-18}$\,erg\,cm$^{-2}$\,s$^{-1}$ in the 0.5--2\,keV band), particularly at $z<4$\cite{Vito2018}.\ Recent constraints were placed on the soft (0.5--2\,keV) X-ray log(N)--log(S) relationship at $z>3.5-5$ using the {\sl Hyper Suprime-Cam/XMM-XXL northern field} sample\cite{pouliasis2022} ($\sim20$\,deg$^{-2}$), but of the AGN with spectroscopic redshifts, only twenty-eight were at $z>3.5$, nine at $z\geq$4.5, and one at $z\geq$5. With its 0.25\,deg$^2$ field-of-view (FOV) and sensitivity down to $3\times10^{-16}$ erg cm$^{-2}$ s$^{-1}$ in 250\,ks (see below), \lem\ will be aptly suited to filling out the gaps in redshift discovery space for high-$z$ AGN.

To determine the number of AGN detectable by \LEM, we must account for the limiting flux as a function of exposure time.\ We first assumed a power law continuum with results shown in {\bf Figure \ref{fig:fluxsensitivity}}.\ The flux limits (red curves) are derived by simulating a grid of sources with a known range of observed 0.5--2 keV fluxes within the \lem\ FOV and running {\sc wavdetect} to determine the detectable sources and hence the detectable fluxes within each exposure. 
The instrumental background and foreground emission is included in the simulated \lem\ fields. 

The number of high-redshift AGN detectable across the \lem\ nominal mission lifetime can then be estimated through two means: by comparing \lem’s sensitivity limits and area coverage to (1) the known AGN log(N)--log(S) estimates as a function of redshift\cite{Vito2018,pouliasis2022} or for the general AGN population regardless of redshift\cite{Lehmer2012}, or (2) the known AGN co-moving volume density as a function of redshift\cite{Kalfountzou2014}.\ Based on the AGN log(N)--log(S) relationship derived from the 4\,Ms {\sl Chandra Deep Field South} (CDF-S)\cite{Lehmer2012}, {\bf \lem\ is expected to detect on the order of $\mathbf{\gtrsim150,000}$ AGN during the nominal mission lifetime.}\ While the vast majority of these AGN reside at lower redshift ($z<3$), \lem\ is expected to detect $\approx6900$, $\approx3400$, and $\approx500$ AGN at redshift $z>3.5$, $z>4$, and $z>5$ when considering the redshift dependent log(N)--log(S) distributions\cite{Vito2018,pouliasis2022}. We note that log(N)--log(S) estimations for $z>5$ are still highly uncertain\cite{Vito2018,pouliasis2022}, and the prediction above may be an underestimate as we discuss below. 

\begin{figure*}[ht]
    \includegraphics[width=0.9\textwidth]{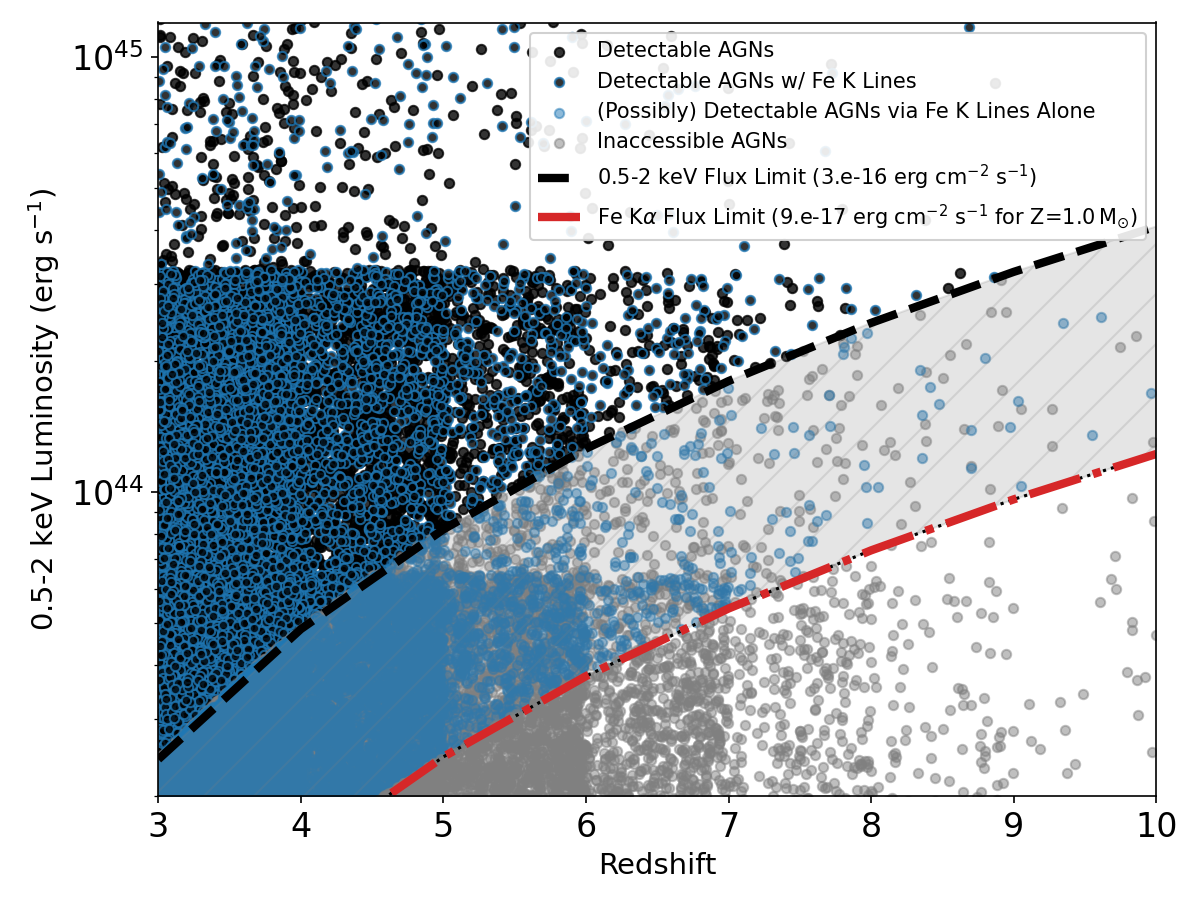}
    \caption{\footnotesize{{\bf Expected \lem-detected AGN luminosity distributions as a function of redshift} for 50\,deg$^2$ with 250\,ks depth per exposure across the \lem\ nominal mission. The dashed black line indicates a continuum source flux sensitivity limit of $3\times10^{-16}$ erg cm${^{-2}}$ s${^{-1}}$
    in the 0.5--2\,keV band while the red dash-dotted line indicates an Fe K$\alpha$ line flux sensitivity limit of $9\times10^{-17}$\,erg\,cm${^{-2}}$\,s${^{-1}}$ in the 0.5--2\,keV band (assuming solar abundances). Black points indicate AGN detectable by \lem\ above the continuum source limits while non-detectable AGN are shown in gray.  Black points with blue outlines and blue points are AGN that may have detectable Fe K$\alpha$ line emission ($\sim40$\% of the total population, again assuming solar abundances); blue points between the black and red curves represent AGN that may be detectable by \lem\ due to the {\it Fe\,K$\alpha$ lines alone}. Dashed gray regions similarly represent parameter spaces where AGN could be detected via Fe\,K$\alpha$ lines, but not as continuum sources. The coarseness of the AGN distributions is due to the adopted redshift and luminosity bins.
  }} 
    \label{fig:highzdistr}
\end{figure*}

We can derive more refined high-$z$ AGN number estimates across specific redshift ranges by using the known AGN co-moving volume density as a function of redshift\cite{Kalfountzou2014}, which has been derived for three different luminosity bins: 43.4 < log($L_X$/erg\,s${^{-1}}$) $<$ 44.0, 44.0 $<$ log($L_X$/erg\,s${^{-1}}$) $<$ 44.7, and log($L_X$/erg\,s${^{-1}}$) $>$ 44.7, where $L_X$ is the 2\--10\,keV luminosity.\ Across the \lem\ nominal mission lifetime, \lem\  is expected to detect $>$14000, $>$2600, and $>$650 X-ray AGN in the redshift bins 3--4, 4--5, and 5--6, respectively (again, down to a limiting flux of 3$\times$10$^{-16}$\,erg\,cm$^{-2}$\,s$^{-1}$ in the 0.5 to 2\,keV band).\ We show these AGN mock distributions in {\bf Figure~\ref{fig:highzdistr}}, where our AGN detected by \lem\ are shown as black points.\ We also plot the limiting flux depths for continuum point sources (3$\times$10$^{-16}$\,erg\,cm$^{-2}$\,s$^{-1}$, black dashed line).\ Grey points are AGN that are not detectable.\ Other curves are defined below.

The very high redshift ($z>6$) AGN luminosity function is already being constrained at longer wavelengths (e.g., the optical and/or near-IR)\cite{greene2024}, and we are only beginning to detect candidate X-ray counterparts\cite{bogdan2024,kovacs2024} for such high redshift AGNs. \lem\ will provide some of the first robust constraints for the X-ray AGN log(N)--log(S) distribution and the X-ray AGN co-moving volume densities for the redshift bins 7--8, 8--9, and 9--10, where we expect \lem\ to detect on average $\sim$200, $\sim$50, and $\sim$10 AGN, respectively, across the nominal mission lifetime ({\bf Figure~\ref{fig:highzdistr}}). The detectable AGN estimates for $z=4-10$ derived via the AGN co-moving volume generally agree well with the estimates provided from log(N)--log(S) for $z>4$ above, though our estimates for $z>5$ are subject to the intrinsic uncertainties in these relations. If the co-moving volume estimates offer a more realistic prediction, then the log(N)--log(S) estimates above for $z>5$ are underestimated by a factor of $\sim2.5$. 

We note that the estimates calculated above deal strictly with the detectability of continuum point sources and will generally require multi-wavelength follow-up.

\subsection{Fe K$\alpha$ lines at high redshift}
\label{highzfek}

\begin{figure}[h]
    \centering
    \includegraphics[width=0.48\textwidth]{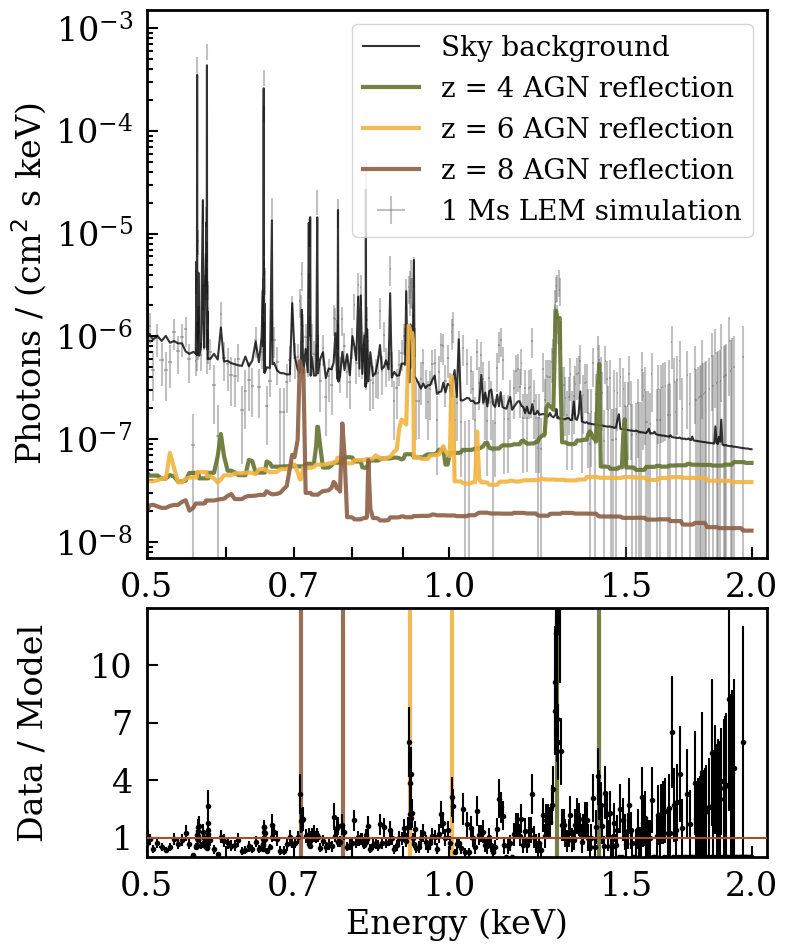}
    \caption{\footnotesize{{\bf High-redshift AGN are separated spectrally in one long exposure.} This is a 1\,Ms \lem\ simulation of the sky background (foreground and instrumental) and three AGN. Each AGN is represented with a power-law with $\Gamma$=1.8 at z=4 (6, 8) absorbed with ($N_{\rm{H}}\gtrsim$10$^{24}$ cm$^{-2}$), 5\% of a scattered unabsorbed component, and {\sc borus} reflection.\ The AGN are normalized so that the intrinsic $L_X(2-10)={10^{45}}$ erg s${^{-1}}$. Vertical lines mark the expected energies of the Fe\,K$\alpha$ and Fe\,K$\beta$ lines in the observed frame. The upper panel shows the sky background and reflection components from the three AGN. The lower panel shows the data/model ratio for data fitted with the sky background model only.}}
    \label{fig:fekspectra}
\end{figure}

A substantial fraction of these AGN will also present detectable Fe K$\alpha$ emission that arises from the accretion disk, the surrounding obscuring torus, the BLR, or a combination of all.\ It is expected that a significant fraction ($\sim$40--50\%\cite{buchner2015,ananna2019,carroll2023}) of AGN at higher redshift (and across cosmic time\cite{buchner2015,lanzuisi2018}) should be Compton-thick ($N_{\rm{H}}\gtrsim$10$^{24}$\,cm$^{-2}$), and thus we expect that $\gtrsim$40\% of the AGN detectable by \lem\ should exhibit prominent Fe\,K$\alpha$ lines.\ We simulate how \lem\ could spectrally separate multiple high-redshift AGN in one long exposure ({\bf Figure~\ref{fig:fekspectra}}).

In {\bf Figure~\ref{fig:fluxsensitivity}} (purple curve, Section \ref{subsec:src_detections}) we simulated Fe\,K$\alpha$ emitting sources, without a continuum source, once again in a grid of varying flux levels within a \LEM\ FOV and running {\sc wavdetect} to determine the limiting line flux.\ In this latter case, the sensitivity limit for the emission lines is determined relative to a narrow 100\,eV spectral window.\ As before, the instrumental background and foreground emission are included in the simulated \lem\ fields.

Iron abundances are expected to correlate inversely with increasing redshift, and that will affect the strength and detectability of Fe K lines in higher redshift AGN.\ To provide a simple examination of this effect, we simulated a few cases of AGN at $z=5$ across a small range of observed 0.5--2\,keV fluxes for abundance levels of 1.0 (solar) and 0.3. For case\,1, we assumed the AGN was reflection-dominated (no power-law continuum included) and simulated this reflection component using the \textsc{borus} model\cite{Balokovic2018}.\ For case\,2, we assumed the AGN could be modeled using an intrinsic power-law continuum and the {\sc borus} reflection component.\ In both of these cases, we included foreground emission and the instrumental background (both generated in \textsc{SOXS}\cite{zuhone2023}), and probed the detectability of Fe K$\alpha$ lines for abundance levels of 1.0 and 0.3 for observed 0.5--2\,keV fluxes of $1\times10^{-15}$, $3\times10^{-16}$ (the standard flux limit in a 250\,ks exposure), and $9\times10^{-17}$\,erg\,cm$^{-2}$\,s$^{-1}$ (the expected limiting flux for Fe\,K lines at Solar abundance levels in a 250\,ks exposure). 

To test for the detectability of an Fe\,K$\alpha$ line for a given model, abundance level, and flux level, we fit each case using a simple power-law (to account for the continuum, whether it be AGN or background-driven) and examined whether the addition of a Gaussian line at the line position resulted in a statistically improved fit (here we used Cash\cite{Cash1979} statistics and compared the $\Delta$C-stat value).\ This type of analysis could result in a full paper in its own right, and we therefore leave further analysis to future works, but the general result is that the Fe\,K$\alpha$ line (naturally) becomes increasingly difficult to detect as one pushes to lower abundance values (0.3) and/or lower flux levels.\ For example, for the \textsc{borus} and \textsc{borus}+{\sc power-law} case, the Fe\,K$\alpha$ can be detected in a spectrum with a 0.5--2\,keV flux of $3\times10^{-16}$\,erg\,cm$^{-2}$\,s$^{-1}$ for abundances of 1.0 at the limiting flux level of $9\times10^{-17}$\,erg\,cm$^{-2}$\,s$^{-1}$ but the line is too faint for an abundance of 0.3.

In {\bf Figure~\ref{fig:highzdistr}}, for our mock distributions of AGNs as a function of redshift, we plot the Fe\,K$\alpha$ limiting line flux ($9\times10^{-17}$\,erg\,cm${^{-2}}$\,s${^{-1}}$, red dash-dotted); this curve assumes solar abundances.\ Black points encased in blue circles represent the $\sim$40\% of sources expected to exhibit prominent Fe\,K$\alpha$ lines (see below); blue points that sit between the black and red curves are AGN that would not be detected as continuum sources but could be detected as Fe\,K$\alpha$ emitters if the line fluxes are in excess of the limiting line flux.

Taking this into consideration, we may expect to detect on the order of $\lesssim$3000, $\lesssim$1100, $\lesssim$300 Fe K$\alpha$ lines (down to the flux limit of $9\times10^{-17}$\,erg\,s$^{-1}$) at redshifts 3--4, 4--5, and 5--6 across the nominal mission.\ The exact number of detectable lines will be a strong function of Fe abundance at a particular redshift.\ These Fe K$\alpha$ lines will provide direct spectroscopic constraints on the redshift distributions of these objects without necessarily the need for follow-up spectroscopic observations to determine the redshift.\ With the sensitivity of \lem\ to emission line features, narrow energy ranges can be efficiently searched for redshifted Fe\,K lines down to a limiting line flux of $9\times10^{-17}$ erg cm${^{-2}}$ s${^{-1}}$ (for solar abundances) in a 250\,ks exposure.\ Beyond $z=5$, where abundance levels may drop to 0.1 solar or lower, these lines may not be prevalent in AGN at all, even Compton-thick AGN.\ We cannot expect to find a large number of Fe K$\alpha$ lines beyond $z=5-6$ based on this simple analysis.\ But with its high spectral resolution that lets us easily pick out even faint lines above the background, \lem\ will provide some of the first observational constraints on the prevalence of Fe K$\alpha$ lines at higher redshifts.

\subsection{Stacking photons}
\label{stacking}

The comprehensive \lem\ All-Sky Survey of 16 Msec will provide even more source photons from faint background sources that can be added together and studied.\ This serendipitous science provides a further rich opportunity for AGN studies. 

Following previous studies\cite{Iwasawa2012,Corral2011}, spectra can be stacked according to AGN type and properties such as column densities, accretion rates, Eddington ratios, and the evolution of the relationship between X-ray reflection strength and the intrinsic AGN source luminosity can be probed.\ \textit{JWST} has now provided a detailed study of AGN hosts from z$\sim3-5$ \cite{Yang2023}.\ \lem~is perfectly suited to provide complementary observations and high-energy X-ray insight into this population of host galaxies, allowing for more detailed study of the accretion properties of AGN that reside inside them.\ The {\sl COSMOS-Web}\cite{casey2022}, {\sl Roman Space Telescope} and {\sl Euclid}\cite{euclid19,marshall2020} samples are excellent candidates for \lem~stacking studies.

\begin{figure}[ht]
    \centering
    \includegraphics[width=0.5\textwidth]{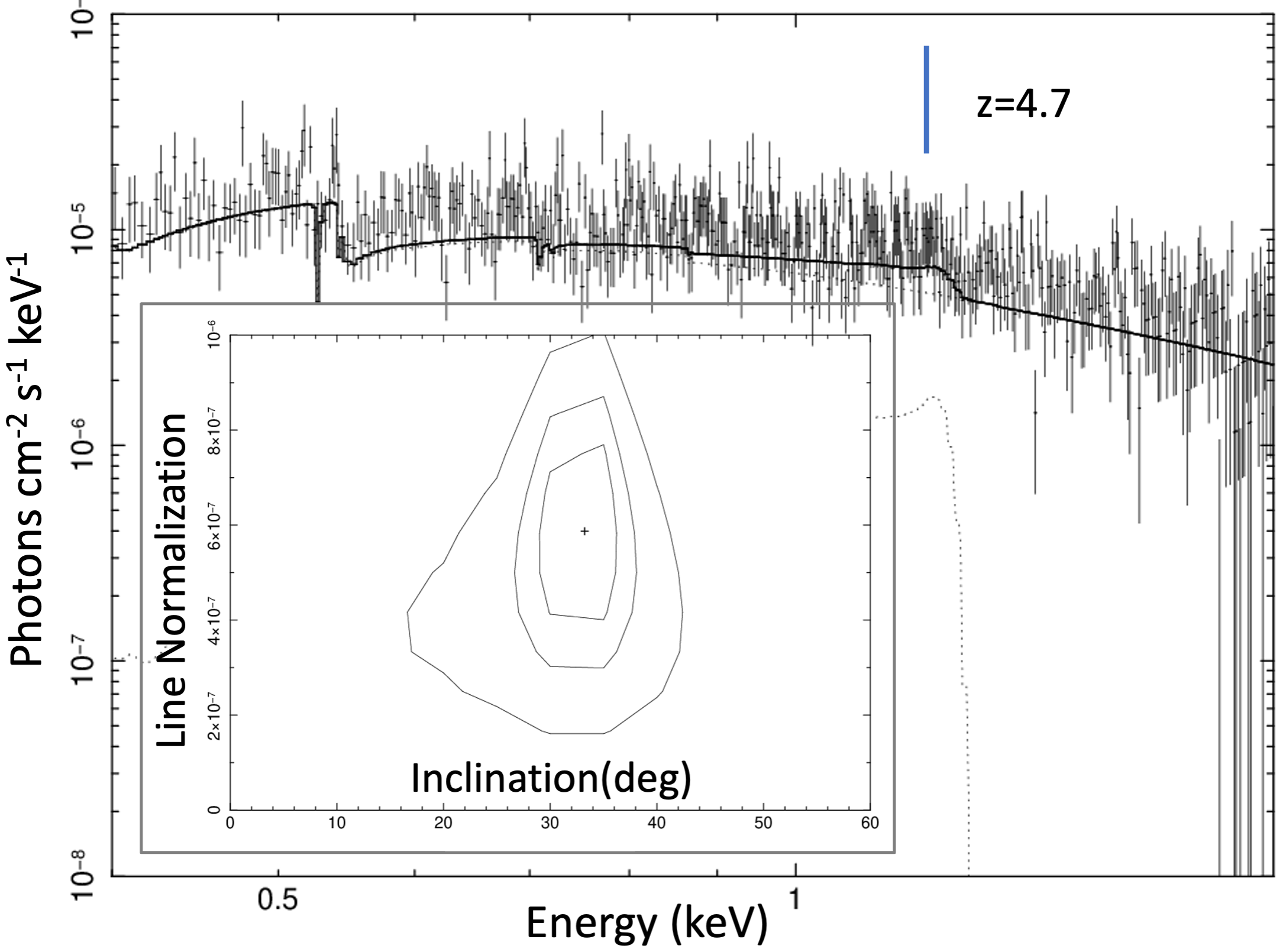}
    \caption{\footnotesize{{\bf \lem\ will detect broad Fe K$\alpha$ lines from accretion disks in quasars.} A 200~ks \LEM\ simulation of a quasar at $z=4.7$ including a relativistic {\sc laor} line profile to model the accretion disk emission.} 
    }
    \label{fig:qso}
\end{figure}

\subsection{Fe\,K line diagnostics for AGN at medium to high redshift}
\label{Fekdiagnostics}

From {\sl XMM-COSMOS}, it was found that the stacked spectra of Type\,1 AGN with L$_{X}$ $\sim10^{44}$ erg s$^{-1}$ typically posess an Fe K$\alpha$ emission line at 6.4\,keV plus high-ionization lines of Fe\,XXV and Fe\,XXVI\cite{Iwasawa2012}. The high-ionization Fe\,K lines appear to have a connection with high accretion rates and these may become stronger with redshift relative to Fe K$\alpha$.\ A \lem\ simulation for 200\,ks at z=4 for  $L_{\rm 2-10\,keV}=10^{44}$\,erg\,s$^{-1}$ is incredibly sensitive to narrow lines at 6.7\,keV and 6.97\,keV with an equivalent width (EW) of 30\,eV or greater.\ \lem\ discovery space will include these high ionization Fe\,K lines.\ With these lines, we can probe high accretion rates and high Eddington ratios for individual AGN, and thus begin statistically robust studies for this recently discovered source population.

\lem\ is also sensitive to details of the accretion disk for AGN at medium redshift, assuming optimistic abundances.\ At typical 0.5 to 7.0 keV fluxes for young radio quasars of a few times 10$^{-14}$\,erg\,cm$^{-2}$\,s$^{-1}$\cite{Snios2020}, \lem\ can constrain the disk inclination.\ We performed a 250\,ks simulation that would correspond to a deep PI-led planned \lem\ pointed observation field.\ To model a relativistic disk line in a basic fashion we used the {\sc laor} model in {\sc xspec} \cite{Laor1991} and fixed the inner and outer disk radii to be 1.23 and 400\,$R_{\rm g}$, respectively ({\bf Figure~\ref{fig:qso}}).\ This simulated source is designed to approximate the Chandra detections of quasars at 4.5$<z<$5.0 \cite{Snios2020}.\ The disk inclination is very well constrained, as is the assumed EW of 70\,eV. A known redshift is required as \lem\ cannot easily constrain redshifts from faint, very broad lines. We identify this science as either using archival data from the PI-led deep pointings, all-sky survey data to follow-up on known quasars, or for GO science to focus on specific targets. 

\section{Other science themes}

\begin{figure}[ht]
    \centering
    \includegraphics[width=0.482\textwidth]{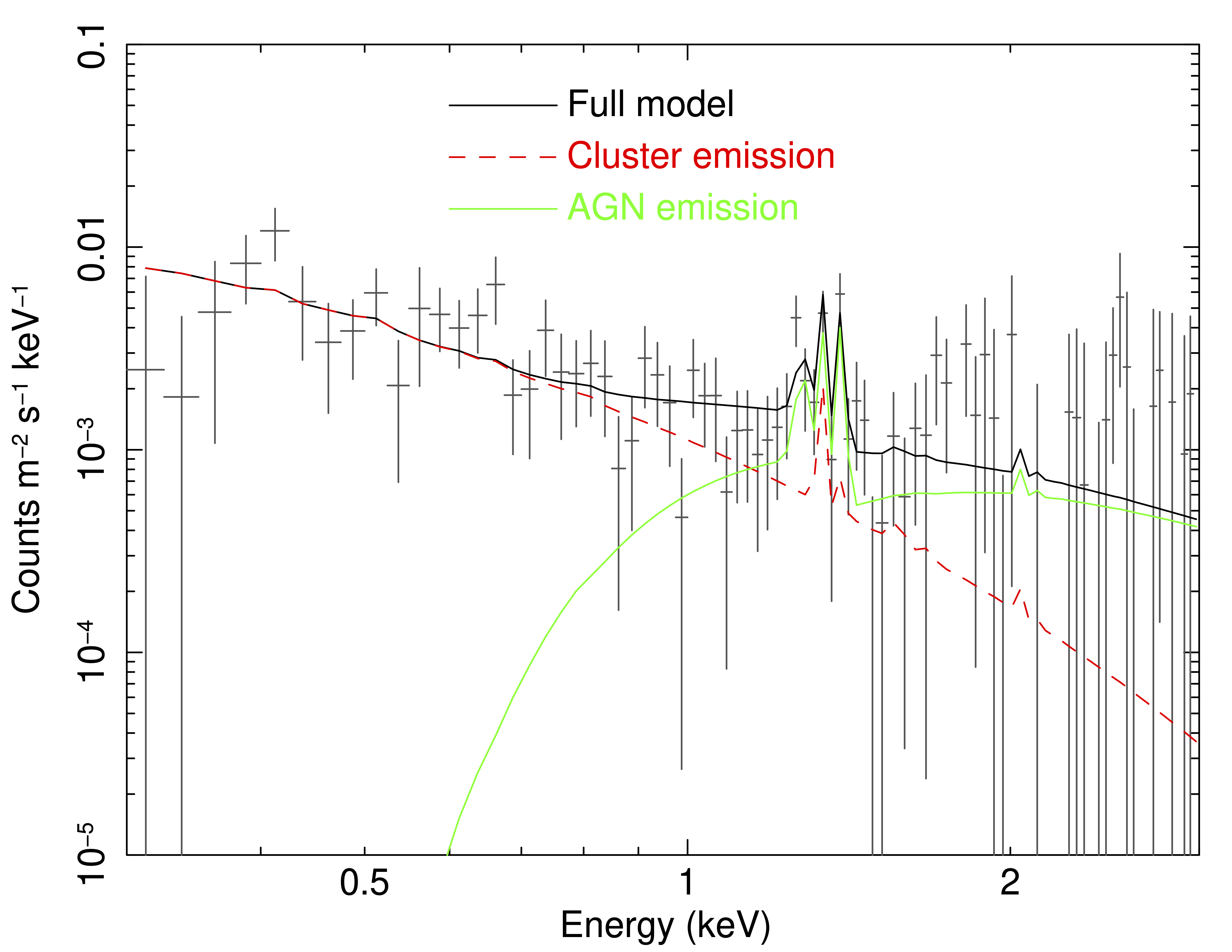}
    \caption{\footnotesize{{\bf Deep \lem\ observations of protoclusters will provide serendipitous AGN science.} A \lem\ simulation for 500~ks of a hypothetical protocluster and the embedded Compton-thick AGN at $z$=4 with L$_X$(2--10\,keV) = 1$\times$10$^{44}$\,erg\,s${^{-1}}$, detected at $>3\sigma$ significance.} 
    }
    \label{fig:proto}
\end{figure}

\subsection{AGN in protoclusters}
\label{proto}

In the next decade, we anticipate the emergence of groundbreaking insights into the first epoch of galaxy cluster assembly at $z\sim$1.5-4 \citep{Chiang2013} by the discovery of statistically significant samples of galaxy overdensities.\ Part of these will be the progenitor groups that will merge to build the massive galaxy clusters that we observe today \citep{Remus2023} and that we call ``protoclusters.''
Because AGN are bright, they act as lighthouses to discover galaxy overdensities, protoclusters, and clusters at high redshift \citep{14cast, 14hatch, 17daddi, 17pat, Mei2023}.\ While the space mission {\sl Euclid} will detect thousands of high-redshift protoclusters in the infrared, it will be highly incomplete at redshift $z>2$, limiting the discovery potential during these critical epochs of cluster assembly. 

\lem\ will open a new window not only to detect high-$z$ protoclusters but especially to study in detail their physical properties unlike any other observatory and answer key questions on the nature of their intergalactic medium and AGN feedback.\ \textit{Herschel/SPIRE} studies indicate an AGN fraction of $0.1 - 0.2$ in protocluster candidates, with higher AGN fractions in denser environments \citep{gatica2024}.\ For galaxy protoclusters at redshift $\sim3$, this AGN fraction appears to increases by a factor of up to six for the highest density regions \citep{lehmer2009a,lehmer2009b}.\ Some AGN in protoclusters are Compton-thick and would be expected to show strong Fe\,K$\alpha$ features \citep{gilli2019}.\ These may be detectable by \lem\ at $z\gtrsim3$ as the lines will be redshifted into the \textit{LEM} passband (see Section~\ref{sec:agnhighz}). 

One of the \lem\  science goals is to observe and study the intergalactic medium in protoclusters, and deep observations of these sources will provide additional serendipitous AGN science. We have performed a 500\,ks \lem\ simulation of a Compton-thick AGN with L$_X$(2--10\,keV) = 1$\times$10$^{44}$\,erg\,s${^{-1}}$, $\Gamma$=1.8, and Fe\,K$\alpha$ and Fe\,K$\beta$ Gaussian emission lines, adding a standard cluster model with L$_{\rm bol} = 1\times10^{44}$ erg s${^{-1}}$, $kT$=5\,keV, and an assumption of Z=0.3\,Z$_\odot$, drawn from observations of ICM enrichment at $z=4$\cite{Mernier2023}. Both the AGN and the surrounding protocluster were placed at $z=4$ ({\bf Figure~\ref{fig:proto}}). The iron lines from the AGN are clearly separated from the Fe\,K 6.7\,keV and 6.97\,keV (restframe energy) lines from the cluster inter-galactic medium.\ \lem\ will probe the impact of non-gravitational heating (e.g., from AGN outflows) on these protocluster environments and help to quantify black-hole growth in high-z galaxy overdensities. 

\subsection{\lem-gravitational wave facilities synergies}
\label{gw}

\lem\ will play a key role in multi-messenger astronomy by searching for the X-ray electromagnetic counterparts of gravitational wave sources.\ 
With space-based gravitational wave observatories, such as \textit{LISA}\cite{Amaro-Seoane2023}, a $3\times10^5$ M$_\odot$ inspiralling massive black hole binary at redshift $z=1$  will be localized with a median precision of 10\,deg$^2$ at one week from coalescence \cite{mangiagli2022}.\ This ‘heads-up’ will provide ample time for facilities like the {\sl Vera Rubin Observatory} to pinpoint the source's location with high precision.\ \lem\ could then act as a monitoring X-ray facility to observe the dimming and brightening of thermal emission of a source prior to and following a coalescence event \cite{krauth2023}.\ The X-ray sky is also less crowded than the infrared or optical sky, making the X-rays an extremely useful wavelength when attempting to identify a counterpart; in some cases, \lem\ may be able to act as a follow-up facility, tiling the sky in an attempt to further localize the gravitational wave sources.

\section{Conclusions}

This white paper represents the culmination of a $\sim$two year-long effort from the AGN science working group involved with the \lem\ proposal process.\ As we have shown, \lem\ has the potential to revolutionize our understanding of the AGN phenomenon in the crucial low-energy X-ray bandpass that has been underutilized to date.

In separating observational signatures of photoionized plasmas from thermal plasmas, \lem\ will probe the AGN and host galaxy connections out to high redshifts.\ Black hole accretion and feedback can be measured on small scales via extreme AGN winds and reverberation mapping, while AGN feedback will be probed on large scales via outflows and long-term variability monitoring. 

\lem\ will detect tens of thousands of AGN at $z\ge2.5$ down to a limiting flux of 3$\times$10$^{-16}$\,erg\,cm$^{-2}$\,s$^{-1}$ in the 0.5 to 2\,keV band.\ In addition to AGN population studies, \lem\ will provide some of the first robust constraints for log(N)--log(S) and the AGN co-moving volume densities for the redshift bins 7--8, 8--9, and 9--10 across its mission lifetime. 

The specific power provided by high-resolution spectroscopy will allow \lem\ to uncover detailed Fe K line properties in quasars at z$\sim$4--5 and disentangle AGN in protoclusters.\ Finally, \lem\ will play a key role in multi-messenger astronomy by searching for the X-ray electromagnetic counterparts of gravitational wave sources.

\section{Acknowledgements}

We thank all of the \lem\ team members who participated in our science discussions and AGN working group meetings to make this collaboration possible.

\small


\vspace{-6mm}
\parindent=0cm
\baselineskip=12pt

\begin{thebibliography}{10}

\bibitem[Kraft et al.(2022)]{kraft2022} Kraft, R., Markevitch, M., Kilbourne, C., et al.\ 2022, arXiv:2211.09827. doi:10.48550/arXiv.2211.09827
\bibitem[National Academies of Sciences, Engineering, and Medicine (2021)]{astro2020} National Academies of Sciences, Engineering, and Medicine: Pathways to Discovery in Astronomy and Astrophysics for the 2020s, The National Academies Press, Washington, DC (2021). doi:10.17226/26141"
\bibitem[ZuHone et al.(2024)]{ZuHone2023} ZuHone, J.~A., Schellenberger, G., Ogorza{\l}ek, A., et al.\ 2024, \apj, 967, 49. doi:10.3847/1538-4357/ad36c1
\bibitem[Bogd{\'a}n et al.(2023)]{Bogdan2023} Bogd{\'a}n, {\'A}., Khabibullin, I., Kov{\'a}cs, O.~E., et al.\ 2023, \apj, 953, 42. doi:10.3847/1538-4357/acdeec
\bibitem[Schellenberger et al.(2023)]{Schellenberger2023} Schellenberger, G., Bogd{\'a}n, {\'A}., ZuHone, J.~A., et al.\ 2023, arXiv:2307.01259. doi:10.48550/arXiv.2307.01259
\bibitem[Mernier et al.(2023)]{Mernier2023} Mernier, F., Su, Y., Markevitch, M., et al.\ 2023, arXiv:2310.04499. doi:10.48550/arXiv.2310.04499
\bibitem[Zhang et al.(2024)]{Zhang2023} Zhang, C., Zhuravleva, I., Markevitch, M., et al.\ 2024, \mnras, 530, 4234. doi:10.1093/mnras/stae1022
\bibitem[Fabian(2012)]{2012ARA&A..50..455F} Fabian, A.~C.\ 2012, \araa, 50, 455. doi:10.1146/annurev-astro-081811-125521
\bibitem[King \& Pounds(2015)]{kingpounds15} King, A. \& Pounds, K.\ 2015, \araa, 53, 115. doi:10.1146/annurev-astro-082214-122316
\bibitem{manbelli18} Man, A. \& Belli, S.\ 2018, Nature Astronomy, 2, 695. doi:10.1038/s41550-018-0558-1
\bibitem[Alton et al.(1999)]{Alton99} Alton, P.~B., Davies, J.~I., \& Bianchi, S.\ 1999, \aap, 343, 51
\bibitem[Dermer \& Giebels(2016)]{Dermer16} Dermer, C.~D. \& Giebels, B.\ 2016, Comptes Rendus Physique, 17, 594. doi:10.1016/j.crhy.2016.04.004
\bibitem[G{\'o}mez-Guijarro et al.(2017)]{gomezguijarro17} G{\'o}mez-Guijarro, C., Gonz{\'a}lez-Mart{\'\i}n, O., Ramos Almeida, C., et al.\ 2017, \mnras, 469, 2720. doi:10.1093/mnras/stx1037
\bibitem{kormandyho13} Kormendy, J. \& Ho, L.~C.\ 2013, \araa, 51, 511. doi:10.1146/annurev-astro-082708-101811
\bibitem{kormandyrichstone95} Kormendy, J. \& Richstone, D.\ 1995, \araa, 33, 581. doi:10.1146/annurev.aa.33.090195.003053
\bibitem{mcluredunlop02} McLure, R.~J. \& Dunlop, J.~S.\ 2002, \mnras, 331, 795. doi:10.1046/j.1365-8711.2002.05236.x
\bibitem{sani11} Sani, E., Marconi, A., Hunt, L.~K., et al.\ 2011, \mnras, 413, 1479. doi:10.1111/j.1365-2966.2011.18229.x
\bibitem{ferraresemerrit00} Ferrarese, L. \& Merritt, D.\ 2000, \apjl, 539, L9. doi:10.1086/312838
\bibitem{gebhardt00} Gebhardt, K., Bender, R., Bower, G., et al.\ 2000, \apjl, 539, L13. doi:10.1086/312840
\bibitem{mcconnellma13} McConnell, N.~J. \& Ma, C.-P.\ 2013, \apj, 764, 184. doi:10.1088/0004-637X/764/2/184
\bibitem[Costa et al.(2018)]{costa18} Costa, T., Rosdahl, J., Sijacki, D., et al.\ 2018, \mnras, 479, 2079. doi:10.1093/mnras/sty1514
\bibitem{porquet10} Porquet, D., Dubau, J. \& Grosso, N.\ 2010,  Space Sci Rev 157, 103. https://doi.org/10.1007/s11214-010-9731-2
\bibitem{liedahl99} Liedahl, D.~A., 1999, Lecture Notes in Physics, Berlin Springer Verlag, vol.\ 520, 1999, p.\ 189
\bibitem[Baker \& Menzel(1938)]{bakermenzel83} Baker, J.~G. \& Menzel, D.~H.\ 1938, \apj, 88, 52. doi:10.1086/143959
\bibitem[Ferland(1999)]{ferland99} Ferland, G.~J.\ 1999, \pasp, 111, 1524. doi:10.1086/316466
\bibitem[Luridiana et al.(2009)]{luridiana} Luridiana, V., Sim{\'o}n-D{\'\i}az, S., Cervi{\~n}o, M., et al.\ 2009, \apj, 691, 1712. doi:10.1088/0004-637X/691/2/1712
\bibitem[Chakraborty et al.(2021)]{chakraborty21} Chakraborty, P., Ferland, G.~J., Chatzikos, M., et al.\ 2021, \apj, 912, 26. doi:10.3847/1538-4357/abed4a
\bibitem[Chakraborty et al.(2020)]{chakraborty20c} Chakraborty, P., Ferland, G.~J., Chatzikos, M., et al.\ 2020, \apj, 901, 69. doi:10.3847/1538-4357/abaaac
\bibitem[Chakraborty et al.(2022)]{chakraborty22} Chakraborty, P., Ferland, G.~J., Chatzikos, M., et al.\ 2022, \apj, 935, 70. doi:10.3847/1538-4357/ac7eb9
\bibitem[Chakraborty et al.(2020)]{Chakraborty20a} Chakraborty, P., Ferland, G.~J., Bianchi, S., et al.\ 2020, Research Notes of the American Astronomical Society, 4, 184. doi:10.3847/2515-5172/abc1dd
\bibitem[Chatzikos et al.(2023)]{Chatzikos23} Chatzikos, M., Bianchi, S., Camilloni, F., et al.\ 2023, arXiv:2308.06396. doi:10.48550/arXiv.2308.06396
\bibitem{heckmanbest14} Heckman, T.~M. \& Best, P.~N.\ 2014, \araa, 52, 589. doi:10.1146/annurev-astro-081913-035722
\bibitem{guainazzibianchi07} Guainazzi, M. \& Bianchi, S.\ 2007, \mnras, 374, 1290. doi:10.1111/j.1365-2966.2006.11229.x
\bibitem{kinkhabwala02} Kinkhabwala, A., Sako, M., Behar, E., et al.\ 2002, \apj, 575, 732. doi:10.1086/341482
\bibitem{bogdan17} Bogd{\'a}n, {\'A}., Kraft, R.~P., Evans, D.~A., et al.\ 2017, \apj, 848, 61. doi:10.3847/1538-4357/aa8c76
\bibitem{grupe08} Grupe, D., Komossa, S., Gallo, L.~C., et al.\ 2008, \apj, 681, 982. doi:10.1086/588213
\bibitem[Guainazzi et al.(2009)]{guainazzi09} Guainazzi, M., Risaliti, G., Nucita, A., et al.\ 2009, \aap, 505, 589. doi:10.1051/0004-6361/200912758
\bibitem[Azadi(2017)]{azadi2017} Azadi, M.\ 2017, Ph.D. Thesis
\bibitem[Ranalli et al.(2003)]{ranalli2003} Ranalli, P., Comastri, A., \& Setti, G.\ 2003, \aap, 399, 39. doi:10.1051/0004-6361:20021600
\bibitem[Fragos et al.(2013)]
{fragos2013} Fragos, T., Lehmer, B., Tremmel, M., et al.\ 2013, \apj, 764, 41. doi:10.1088/0004-637X/764/1/41
\bibitem[Mezcua(2019)]{mezcua2019} Mezcua, M.\ 2019, Nature Astronomy, 3, 6. doi:10.1038/s41550-018-0662-2
\bibitem[Latimer et al.(2021)]{latimer2021} Latimer, L.~J., Reines, A.~E., Bogdan, A., et al.\ 2021, \apjl, 922, L40. doi:10.3847/2041-8213/ac3af6
\bibitem[Sacchi et al.(2024)]{2024arXiv240601707S} Sacchi, A., Bogdan, A., Chadayammuri, U., et al.\ 2024, arXiv:2406.01707. doi:10.48550/arXiv.2406.01707
\bibitem[Bykov et al.(2024)]{bykov2024} Bykov, S.~D., Gilfanov, M.~R., \& Sunyaev, R.~A.\ 2024, \mnras, 527, 1962. doi:10.1093/mnras/stad3355
\bibitem[Condon et al.(1991)]{condon1991} Condon, J.~J., Huang, Z.-P., Yin, Q.~F., et al.\ 1991, \apj, 378, 65. doi:10.1086/170407
\bibitem[Trump et al.(2015)]{trump2015} Trump, J.~R., Sun, M., Zeimann, G.~R., et al.\ 2015, \apj, 811, 26. doi:10.1088/0004-637X/811/1/26
\bibitem[Hainline et al.(2016)]{hainline2016} Hainline, K.~N., Reines, A.~E., Greene, J.~E., et al.\ 2016, \apj, 832, 119. doi:10.3847/0004-637X/832/2/119
\bibitem[Cann et al.(2019)]{cann2019} Cann, J.~M., Satyapal, S., Abel, N.~P., et al.\ 2019, \apjl, 870, L2. doi:10.3847/2041-8213/aaf88d
\bibitem[Satyapal et al.(2021)]{satyapal2021} Satyapal, S., Kamal, L., Cann, J.~M., et al.\ 2021, \apj, 906, 35. doi:10.3847/1538-4357/abbfaf
\bibitem[Sim et al.(2008)]{sim2008} Sim, S.~A., Long, K.~S., Miller, L., et al.\ 2008, \mnras, 388, 611. doi:10.1111/j.1365-2966.2008.13466.x
\bibitem[Matzeu et al.(2023)]{matzeu2023} Matzeu, G.~A., Brusa, M., Lanzuisi, G., et al.\ 2023, \aap, 670, A182. doi:10.1051/0004-6361/202245036
\bibitem[Hagino et al.(2016)]{hagino2016} Hagino, K., Odaka, H., Done, C., et al.\ 2016, \mnras, 461, 3954. doi:10.1093/mnras/stw1579
\bibitem[Parker et al.(2017)]{Parker2017} Parker, M.~L., Pinto, C., Fabian, A.~C., et al.\ 2017, \nat, 543, 83. doi:10.1038/nature21385
\bibitem[Reeves et al.(2018)]{reeves2018} Reeves, J.~N., Braito, V., Nardini, E., et al.\ 2018, \apjl, 854, L8. doi:10.3847/2041-8213/aaaae1
\bibitem[Pinto et al.(2018)]{Pinto2018} Pinto, C., Alston, W., Parker, M.~L., et al.\ 2018, \mnras, 476, 1021. doi:10.1093/mnras/sty231
\bibitem[Xu et al.(2022)]{xu2022} Xu, Y., Pinto, C., Kara, E., et al.\ 2022, \mnras, 513, 1910. doi:10.1093/mnras/stac1058
\bibitem[Nicastro et al.(1999)]{Nicastro1999} Nicastro, F., Fiore, F., Perola, G.~C., \& {Elvis}, M. 1999, \apj, 512, 184. doi:10.1086/306736
\bibitem[Kaastra et al.(2012)]{Kaastra2012} Kaastra, J.~S., Detmers, R.~G., Mehdipour, M., et al.\ 2012, \aap, 539, A117. doi:10.1051/0004-6361/201118161
\bibitem[Rogantini et al.(2022)]{Rogantini22b} Rogantini, D., Mehdipour, M., Kaastra, J., et al.\ 2022, \apj, 940, 122. doi:10.3847/1538-4357/ac9c01
\bibitem[Reeves et al.(2021)]{reeves2021} Reeves, J.~N., Porquet, D., Braito, V., et al.\ 2021, \aap, 649, L3. doi:10.1051/0004-6361/202140953
\bibitem[Kaastra et al.(2014)]{Kaastra2014} Kaastra, J.~S., Kriss, G.~A., Cappi, M., Mehdipour, M., et al.\ 2014, Science, 345, 64. doi:10.1126/science.1253787
\bibitem[Kara et al.(2021)]{Kara2021} Kara, E., Mehdipour, M., Kriss, G.~A., et al.\ 2021, \apj, 922, 151. doi:10.3847/1538-4357/ac2159
\bibitem[Mehdipour et al.(2024)]{Mehdipour24} Mehdipour, M., Kriss, G.~A., Kaastra, J.~S., et al.\ 2024, \apj, 962, 155. doi:10.3847/1538-4357/ad1bcb
\bibitem[Risaliti et al.(2011)]{Risaliti11} Risaliti, G., Nardini, E., Salvati, M., et al.\ 2011, \mnras, 410, 1027. doi:10.1111/j.1365-2966.2010.17503.x
\bibitem[Wang et al.(2022)]{Wang22} Wang, Y., Kaastra, J., Mehdipour, M., et al.\ 2022, \aap, 657, A77. doi:10.1051/0004-6361/202141599
\bibitem[Mehdipour et al.(2022)]{Mehdipour22} Mehdipour, M., Kriss, G.~A., Costantini, E., et al.\ 2022, \apjl, 934, L24. doi:10.3847/2041-8213/ac822f
\bibitem[Kriss et al.(2019)]{Kriss19} Kriss, G.~A., De Rosa, G., Ely, J., et al.\ 2019, \apj, 881, 153. doi:10.3847/1538-4357/ab3049
\bibitem[Parker et al.(2019)]{Parker19} Parker, M.~L., Longinotti, A.~L., Schartel, N., et al.\ 2019, \mnras, 490, 683. doi:10.1093/mnras/stz2566
\bibitem[Rupke(2018)]{Rupke18} Rupke, D.\ 2018, Galaxies, 6, 138. doi:10.3390/galaxies6040138
\bibitem[Veilleux et al.(2005)]{Veilleux2005} Veilleux, S., Cecil, G., \& Bland-Hawthorn, J.\ 2005, \araa, 43, 769. doi:10.1146/annurev.astro.43.072103.150610
\bibitem[HeckmanBest (2023)]{heckman2023} Heckman, T.~M.\ \& Best, P.~N.\ 2023, {\it Galaxies}, 11, 1. doi:10.3390/galaxies11010021
\bibitem[Tanner et al. (2024)]{tanner2024} Tanner, R., Weaver, K., et al.\ 2024 in prep.
\bibitem[Peterson et al. (2004)]{peterson2004} Peterson, B.~M., Ferrarese, L., Gilbert, K.~M., et al., 2004, \apj, 613, 682, doi:10.1086/423269
\bibitem[Longinotti et al.(2008)]{longinotti2008} Longinotti, A.~L., Nucita, A., Santos-Lleo, M., et al.\ 2008, \aap, 484, 311. doi:10.1051/0004-6361:200809374
\bibitem[Grier et al.(2012)]{Grier2012} Grier, C.~J., Peterson, B.~M., Pogge, R.~W., et al.\ 2012, \apjl, 744, L4. doi:10.1088/2041-8205/744/1/L4
\bibitem[Mullaney \& Ward(2008)]{mullanyward2008} Mullaney, J.~R. \& Ward, M.~J.\ 2008, \mnras, 385, 53. doi:10.1111/j.1365-2966.2007.12777.x
\bibitem[Tremou et al.(2015)]{tremou2015} Tremou, E., Garcia-Marin, M., Zuther, J., et al.\ 2015, \aap, 580, A113. doi:10.1051/0004-6361/201525707
\bibitem[Grier et al.(2017)]{grier2017} Grier, C.~J., Trump, J.~R., Shen, Y., et al.\ 2017, \apj, 851, 21. doi:10.3847/1538-4357/aa98dc
\bibitem[Guo et al.(2022)]{guo2022} Guo, H., Barth, A.~J., \& Wang, S.\ 2022, \apj, 940, 20. doi:10.3847/1538-4357/ac96ec
\bibitem[U et al.(2022)]{u2022} U, V., Barth, A.~J., Vogler, H.~A., et al.\ 2022, \apj, 925, 52. doi:10.3847/1538-4357/ac3d26
\bibitem[Homayouni et al.(2022)]{homayouni2022} Homayouni, Y., Sturm, M.~R., Trump, J.~R., et al.\ 2022, \apj, 926, 225. doi:10.3847/1538-4357/ac478b
\bibitem[Bentz \& Katz(2015)]{bentz2015} Bentz, M.~C. \& Katz, S.\ 2015, \pasp, 127, 67. doi:10.1086/679601
\bibitem[Kova{\v{c}}evi{\'c} et al.(2022)]{kovacevic2022} Kova{\v{c}}evi{\'c}, M., Pasquato, M., Marelli, M., et al.\ 2022, \aap, 659, A66. doi:10.1051/0004-6361/202142444
\bibitem[Venturi et al. (2018)]{venturi2018} Venturi, G., Nardini, E., Marconi, A., et al., 2018, \aap, 619, 74V, doi:10.1051/0004-6361/201833668
\bibitem[Middei et al.(2017)]{middei2017} Middei, R., Vagnetti, F., Bianchi, S., et al.\ 2017, \aap, 599, A82. doi:10.1051/0004-6361/201629940
\bibitem[Khabibullin et al.(20323)]{Khabibullin2023} Khabibullin, I., Galeazzi, M., et al.\ 2023, arXiv:2310.16038. doi:10.48550/arXiv.2310.16038
\bibitem[Finkelstein et al.(2022)]{finkelstein2022} Finkelstein, S.~L., Bagley, M.~B., Haro, P.~A., et al.\ 2022, \apjl, 940, L55. doi:10.3847/2041-8213/ac966e
\bibitem[Kocevski et al.(2023)]{kocevski2023} Kocevski, D.~D., Barro, G., McGrath, E.~J., et al.\ 2023, \apjl, 946, L14. doi:10.3847/2041-8213/acad00
\bibitem[Yang et al.(2023)]{Yang2023} Yang, J., Wang, F., Fan, X., et al.\ 2023, \apjl, 951, L5. doi:10.3847/2041-8213/acc9c8
\bibitem[Casey et al.(2022)]{casey2022} Casey, C.~M., Kartaltepe, J.~S., Drakos, N.~E., et al.\ 2022, arXiv:2211.07865. doi:10.48550/arXiv.2211.07865
\bibitem[Marshall et al.(2022)]{Marshall2022} Marshall, M.~A., Watts, K., Wilkins, S., et al.\ 2022, \mnras, 516, 1047. doi:10.1093/mnras/stac2111
\bibitem[Euclid Collaboration et al.(2019)]{euclid19} Euclid Collaboration, Barnett, R., Warren, S.~J., et al.\ 2019, \aap, 631, A85. doi:10.1051/0004-6361/201936427
\bibitem[Marshall et al.(2020)]{marshall2020} Marshall, M.~A., Ni, Y., Di Matteo, T., et al.\ 2020, \mnras, 499, 3819. doi:10.1093/mnras/staa2982
\bibitem[Vito et al.(2014)]{Vito2014} Vito, F., Gilli, R., Vignali, C., et al.\ 2014, \mnras, 445, 3557. doi:10.1093/mnras/stu2004
\bibitem[Kalfountzou et al.(2014)]{Kalfountzou2014} Kalfountzou, E., Civano, F., Elvis, M., et al.\ 2014, \mnras, 445, 1430. doi:10.1093/mnras/stu1745
\bibitem[Georgakakis et al.(2015)]{Georgakakis2015} Georgakakis, A., Aird, J., Buchner, J., et al.\ 2015, \mnras, 453, 1946. doi:10.1093/mnras/stv1703
\bibitem[Vito et al.(2018)]{Vito2018} Vito, F., Brandt, W.~N., Yang, G., et al.\ 2018, \mnras, 473, 2378. doi:10.1093/mnras/stx2486
\bibitem[\protect\citeauthoryear{Elvis et al.}{2009}]{elvis2009} Elvis M., Civano F., Vignali C., et al., 2009, ApJS, 184, 158. doi:10.1088/0067-0049/184/1/158
\bibitem[\protect\citeauthoryear{Civano et al.}{2012}]{civano2012} Civano F., Elvis M., Brusa M., et al., 2012, ApJS, 201, 30. doi:10.1088/0067-0049/201/2/30
\bibitem[\protect\citeauthoryear{Kim et al.}{2007}]{kim2007} Kim M., Kim D.-W., Wilkes B.~J., et al., 2007, ApJS, 169, 401. doi:10.1086/511634
\bibitem[\protect\citeauthoryear{Green et al.}{2009}]{green2009} Green P.~J., Aldcroft T.~L., Richards G.~T., et al., 2009, ApJ, 690, 644. doi:10.1088/0004-637X/690/1/644
\bibitem[\protect\citeauthoryear{Pouliasis et al.}{2022}]{pouliasis2022} Pouliasis E., Georgantopoulos I., Ruiz A., et al., 2022, A\&A, 658, A175. doi:10.1051/0004-6361/202142059
\bibitem[Lehmer et al.(2012)]{Lehmer2012} Lehmer, B.~D., Xue, Y.~Q., Brandt, W.~N., et al.\ 2012, \apj, 752, 46. doi:10.1088/0004-637X/752/1/46
\bibitem[Greene et al.(2024)]{greene2024} Greene, J.~E., Labbe, I., Goulding, A.~D., et al.\ 2024, \apj, 964, 39. doi:10.3847/1538-4357/ad1e5f
\bibitem[Bogd{\'a}n et al.(2024)]{bogdan2024} Bogd{\'a}n, {\'A}., Goulding, A.~D., Natarajan, P., et al.\ 2024, Nature Astronomy, 8, 126. doi:10.1038/s41550-023-02111-9
\bibitem[Kov{\'a}cs et al.(2024)]{kovacs2024} Kov{\'a}cs, O.~E., Bogd{\'a}n, {\'A}., Natarajan, P., et al.\ 2024, \apjl, 965, L21. doi:10.3847/2041-8213/ad391f
\bibitem[Buchner et al.(2015)]{buchner2015} Buchner, J., Georgakakis, A., Nandra, K., et al.\ 2015, \apj, 802, 89. doi:10.1088/0004-637X/802/2/89
\bibitem[Ananna et al.(2019)]{ananna2019} Ananna, T.~T., Treister, E., Urry, C.~M., et al.\ 2019, \apj, 871, 240. doi:10.3847/1538-4357/aafb77
\bibitem[Carroll et al.(2023)]{carroll2023} Carroll, C.~M., Ananna, T.~T., Hickox, R.~C., et al.\ 2023, \apj, 950, 127. doi:10.3847/1538-4357/acc402
\bibitem[Lanzuisi et al.(2018)]{lanzuisi2018} Lanzuisi, G., Civano, F., Marchesi, S., et al.\ 2018, \mnras, 480, 2578. doi:10.1093/mnras/sty2025
\bibitem[Balokovi{\'c} et al.(2018)]{Balokovic2018} Balokovi{\'c}, M., Brightman, M., Harrison, F.~A., et al.\ 2018, \apj, 854, 42. doi:10.3847/1538-4357/aaa7eb
\bibitem[\protect\citeauthoryear{ZuHone et al.}{2023}]{zuhone2023} ZuHone J.~A., Vikhlinin A., Tremblay G.~R., et al. 2023, ascl.soft. ascl:2301.024
\bibitem[Cash(1979)]{Cash1979} Cash, W.\ 1979, \apj, 228, 939. doi:10.1086/156922
\bibitem[Iwasawa et al.(2012)]{Iwasawa2012} Iwasawa, K., Mainieri, V., Brusa, M., et al.\ 2012, \aap, 537, A86. doi:10.1051/0004-6361/201118203
\bibitem[Corral et al.(2011)]{Corral2011} Corral, A., Della Ceca, R., Caccianiga, A., et al., 2011, \aap, 530, A42. 10.1051/0004-6361/201015227
\bibitem[Snios et al.(2020)]{Snios2020} Snios, B., Siemiginowska, A., Sobolewska, M., et al.\ 2020, \apj, 899, 127. doi:10.3847/1538-4357/aba2ca
\bibitem[Laor(1991)]{Laor1991} Laor, A.\ 1991, \apj, 376, 90. doi:10.1086/170257
\bibitem[Chiang et al.(2013)]{Chiang2013} Chiang, Y.-K., Overzier, R., \& Gebhardt, K.\ 2013, \apj, 779, 127. doi:10.1088/0004-637X/779/2/127
\bibitem[Remus et al.(2023)]{Remus2023} Remus, R.-S., Dolag, K., \& Dannerbauer, H.\ 2023, \apj, 950, 191. doi:10.3847/1538-435/accb91
\bibitem[Castignani et al.(2014)]{14cast}  Castignani, G., Chiaberge, M., Celotti, A., et al.\ 2014, \apj, 792, 114. doi:10.1088/0004-637X/792/2/114
\bibitem[Hatch et al.(2014)]{14hatch} Hatch, N.~A., Wylezalek, D., Kurk, J.~D., et al.\ 2014, \mnras, 445, 280. doi:10.1093/mnras/stu1725
\bibitem[Daddi et al.(2017)]{17daddi} Daddi, E., Jin, S., Strazzullo, V., et al.\ 2017, \apjl, 846, L31. doi:10.3847/2041-8213/aa8808
\bibitem[Paterno-Mahler et al.(2017)]{17pat} Paterno-Mahler, R., Blanton, E.~L., Brodwin, M., et al.\ 2017, \apj, 844, 78. doi:10.3847/1538-4357/aa7b89
\bibitem[Mei et al.(2023)]{Mei2023} Mei, S., Hatch, N.~A., Amodeo, S., et al.\ 2023, \aap, 670, A58. doi:10.1051/0004-6361/202243551
\bibitem[Gatica et al.(2024)]{gatica2024} Gatica, C., Demarco, R., Dole, H., et al.\ 2024, \mnras, 527, 3006. doi:10.1093/mnras/stad3404
\bibitem[Lehmer et al.(2009a)]{lehmer2009a} Lehmer, B.~D., Alexander, D.~M., Geach, J.~E., et al.\ 2009, \apj, 691, 687. doi:10.1088/0004-637X/691/1/687
\bibitem[Lehmer et al.(2009b)]{lehmer2009b} Lehmer, B.~D., Alexander, D.~M., Chapman, S.~C., et al.\ 2009, \mnras, 400, 299. doi:10.1111/j.1365-2966.2009.15449.x
\bibitem[Gilli et al.(2019)]{gilli2019} Gilli, R., Mignoli, M., Peca, A., et al.\ 2019, \aap, 632, A26. doi:10.1051/0004-6361/201936121
\bibitem[Amaro-Seoane et al.(2023)]{Amaro-Seoane2023} Amaro-Seoane, P., Andrews, J., Arca Sedda, M., et al.\ 2023, Living Reviews in Relativity, 26, 2. doi:10.1007/s41114-022-00041-y
\bibitem[Mangiagli et al.(2022)]{mangiagli2022} Mangiagli, A., Caprini, C., Volonteri, M., et al.\ 2022, \prd, 106, 103017. doi:10.1103/PhysRevD.106.103017
\bibitem[Krauth et al.(2023)]{krauth2023} Krauth, L.~M., Davelaar, J., Haiman, Z., et al.\ 2023, \mnras, 526, 5441. doi:10.1093/mnras/stad3095

\end{thebibliography}
\end{document}